\documentclass[twocolumn,secnumarabic,superscriptaddress,amssymb, nobibnotes, aps, prb]{revtex4-2} 
\usepackage{epsfig,amssymb,amsmath,graphicx,subfigure,hyperref  }  
\usepackage{mathrsfs}
\usepackage{color}
\usepackage{multirow}
\usepackage{tabularx}
\usepackage[normalem]{ulem}
\usepackage{physics}
\renewcommand{\matrixel}[3]{{#1}^\dagger{#2}{#3}}

\usepackage{cancel}
\renewcommand{\vec}[1]{\boldsymbol{#1}}
\usepackage{xcolor}

\begin{document}
\title{Wannier Topology and Quadrupole Moments for a generalized Benalcazar-Bernevig-Hughes Model}
\author{Liu Yang} 
\email{liu.yang-2@manchester.ac.uk}
\affiliation{Department of Physics and Astronomy, The University of Manchester, Manchester M13 9PL, United Kingdom }
\author{Alessandro Principi}
\affiliation{Department of Physics and Astronomy, The University of Manchester, Manchester M13 9PL, United Kingdom }
\author{Niels R. Walet}
\affiliation{Department of Physics and Astronomy, The University of Manchester, Manchester M13 9PL, United Kingdom } %

\begin{abstract}
We analyze a special separable and chiral-symmetric model with a quantized quadrupole moment, extending the Benalcazar-Bernevig-Hughes model [Science 357, 61 (2017)]. Using nested-Wilson loop formalism, we give an exact expression for Wannier centers, sector polarizations, and quadrupole moments. These are connected to the winding numbers of the constitutive one-dimensional chains. We prove that these winding numbers can characterize the model's Wannier topology as a $\mathbb{Z}\times\mathbb{Z}$ set. These results clearly show that the quantization of the quadrupole moment can arise without additional spatial symmetry (except for translation symmetry) for the bulk. By switching from the Wannier representation to the Bloch representation, we derive an alternative expression for the bulk quadrupole moment and obtain its exact value. Combining the bulk quadrupole and edge polarizations, we analytically calculate the corner charge in a large square system and make the bulk-boundary correspondence explicit in an edge-consistent gauge. Our work reveals the relationship between zero-energy states at the boundary, charge localization, and the bulk quadrupole of the extended model.
\end{abstract}

\maketitle

\section{Introduction}
The use of the mathematical concept of topology has proven very powerful to describe edge states in systems where the topological invariant inside the material differs from that of the outside~\cite{SSH_1979,Solitons,proof_1D,Haldane1988,QWZmodel,KaneQHSE2005,bulk_edge_XiaoLiang2006,TRSinvariant2007,Kane_Z2,ShouchengQSHE,TI3D_liang2007,TIindices_liang2007,QSHE3D_Roy2009,defect_Jeffrey2010,HasanRMP2010,EdgeDirac2011,topo_class,Exact_edge2018}. This concept has since been generalized to higher-order topological states following the work by Benalcazar, Bernevig, and Hughes (BBH)~\cite{multipole_PRB_2017}. These include corner and hinge states, where a system in $d$ dimension can generate topological edge states in $(d-n)$ dimensions ($n>1$)~\cite{multipole_PRB_2017,multipole_science2017,Zhida2017,HOTI_science2018,HOTinversion2018,HOTISinversion2018,second_order2,HOTprinciple2019,FermiArcs2020,Yafei2020,Changan2020,type2_2020,Exact_localized2021}. In two dimensions, a prototypical example of a system with higher-order topological states is the BBH model, described by a simple Hamiltonian with only nearest-neighbor hopping terms and double mirror symmetry. This has been realized in various experimental setups, such as photonic\cite{photonic2019,photonic2020}, phononic~\cite{phononic2018}, acoustic\cite{Acoustic2020}, and microwave- and electrical-circuit systems\cite{microwave_circuit2018,circuit2018,circuit2019}. The fractional corner charge in a finite system of the BBH model can be characterized by an off-diagonal quadrupole moment calculated through the nested-Wilson loop formalism~\cite{multipole_science2017,multipole_science2017,Wilson2019}, where the quadrupole moment is defined as the multiplication of the Wannier-sector polarizations.

In the papers that started this field~\cite{multipole_science2017,multipole_PRB_2017}, the general role of the symmetries in quantizing the quadrupole moment was discussed. It was shown that additional spatial symmetry is needed to quantize the Wannier-sector polarizations and hence for the quadrupole fractionalization. However, by calculating the quadrupole using many-body operators in real space, it has been shown that chiral symmetry can also protect the quantized quadrupole moment in a disordered BBH model \cite{2020DisEQI}. Even though the definitions of the quadrupole moment in Refs.~\cite{multipole_PRB_2017,multipole_science2017} and Ref.~\cite{2020DisEQI} have not been proven equivalent mathematically, the latter study about a model without additional spatial symmetries leads us to ask the following question: Can we quantize the quadrupole moment and characterize the Wannier band topology through the Wilson formalism in some cases beyond the BBH model without additional spatial symmetry? This is the first question we attempt to answer in this paper.

Recently, a systematic framework to calculate the quadrupole moment and the corner charge using the Wannier functions has been introduced in Ref.~\cite{Quadrupole_Wannier2021}. The researchers have applied it to the BBH model and numerically obtained the same corner charge as was found in the original papers, while the quadrupole moment is gauge dependent. As there is no analytical proof of the connection between the quadrupole moment in the Wannier representation~\cite{localizedWannier1997,Resta2011,RMPVanderbilt2012,EQM2020,EQM2021,Watanabe2019,Watanabe2020,Quadrupole_Wannier2021} and the Wilson-loop formalism, it leads to the question about the relationship between the two and especially when they are equal. Therefore, the second question we will try to address is the following: Can we build an analytic connection between these two definitions of the bulk quadrupole moment and thus give the correct correspondence between the bulk quadrupole moment and the corner charge (bulk-boundary correspondence)?

The BBH model is built from one-dimensional (1D) Su-Schrieffer-Heeger (SSH) chains, with a suitable chosen cross-chain coupling. Here, to answer the two questions we have proposed, we study a generalized two-dimensional (2D) model by extending all 1D SSH chains of the BBH model to general two-band chiral-symmetric chains. We call this model the ``generalized BBH model'', which was first proposed in Refs.~\cite{SecondChiral2019,benalcazar2022chiral}. 

We calculate the Wannier centers and polarizations by finding the exact expression for the Wilson loops along two orthogonal axes in the first Brillouin Zone (FBZ). Applying these results to the analysis of the Wannier bands, we show that their topology is indeed characterized by two winding numbers and analyze the discontinuous displacement of the Wannier centers by continuously changing the parameters. Furthermore, we evaluate a simple expression for the quantized quadrupole moment through the Wilson formalism. We find that the Wannier sector polarizations and quadrupole moment are determined by the fractional parts of the winding numbers of the constitutive chains in the corresponding direction. Thus, we conclude that the quantization of the quadrupole moment of the generalized BBH model does not require additional spatial symmetry, such as mirror or $C_4$ rotation symmetries.

Applying the Fourier transformation to the quadrupole evaluated as the expectation value of the position operator in the Wannier functions of the central unit cell\cite{localizedWannier1997,Resta2011,RMPVanderbilt2012,EQM2020,EQM2021,Watanabe2019,Watanabe2020,Quadrupole_Wannier2021}, we can define a quadrupole invariant in the Bloch representation. We calculate this invariant for the bulk of the generalized BBH model in a specific gauge and prove that its fractional part is equal to the quadrupole moment defined in the Wilson formalism~\cite{multipole_PRB_2017,multipole_science2017}. However, as discussed in Refs~\cite{Watanabe2019,Quadrupole_Wannier2021}, the quadrupole invariant is gauge-dependent, and thus cannot be used to define a gauge-invariant corner charge. We then combine it with the edge polarizations with an edge-consistent gauge mimicking the idea from Ref.~\cite{Quadrupole_Wannier2021}. We find that the corner charge is equal to the fractional value given by the quadrupole moment in the Wilson formalism and obtain the correct bulk-boundary correspondence. These results provide an answer to our second question within the generalized BBH model.

\section{1D TWO-BAND INSULATOR WITH CHIRAL SYMMETRY}
\label{sec:1D}
We first review the model of a 1D two-band insulator with chiral symmetry in order to gain insight into the interplay between this symmetry and topology. The Bloch Hamiltonian of the 1D  model can be written as 
\begin{equation}\label{eq:1D_insulator_def}
    h(k)=\vec{b}(k)\cdot\vec{\sigma}=b^1(k)\sigma_1+b^2(k)\sigma_2,
\end{equation} 
where the vector $\vec{b}(k)$ will be called the ``characteristic vector" and $k\in[0,2\pi]$ is a 1D wavevector. The vanishing of the third component $b^3(k)$ is a consequence of chiral symmetry: $\{h(k),\sigma_3\}=0$. The eigen-energies of $h(k)$ are $\pm|\vec{b}(k)|$. 

To define the electric polarization, one should solve for the eigenvalues of the position operator projected onto the occupied bands. This reduces to the calculation of the Wilson loop~\cite{multipole_PRB_2017,multipole_science2017}, as reviewed in the Supplementary Material (SM)~\cite{SM}. It can be proven that the electric polarization is determined by the Zak-Berry phase as an integral of the Berry connection of the occupied band $\vec{\mathcal{A}}(\vec{b})=i \matrixel{\psi_-}{\partial_{\vec{b}}}{\psi_-}$ in the characteristic vector space along a closed loop $\mathcal{C}:k\in[0,2\pi]\to\vec{b}(k)$ as
\begin{align} \label{eq:polarization_definition}
p=\frac{1}{2\pi}\int_{\mathcal{C}}\,d\vec{b}\cdot \vec{\mathcal{A}}(\vec{b})\, \text{mod}\,1=\frac{\mathcal{N}}{2}\, \text{mod}\,1.
\end{align}
Note that we set the electron charge $\text{e}=1$ throughout this paper. The integer $\mathcal{N}$ is the winding number of $\mathcal{C}$ around the origin. Thus, the energy gap $2|\vec{b}(k)|$ must close when $\mathcal{N}$ changes. The topological phases of this 1D model correspond to the AIII topological class~\cite{A3class2014,topo_class}. We choose the eigenstates of Eq.~(\ref{eq:1D_insulator_def}) to be
\begin{equation}
    \psi_\pm=(b^1-ib^2,\pm|\vec{b}|)^T/\sqrt{2|\vec{b}|^2}.\label{1d_gauge}
\end{equation} 
In this gauge choice, the Zak phase in Eq.~(\ref{eq:polarization_definition}) gives the exact fractional winding number. The exact number of zero modes at each endpoint of a finite chain is also equal to this winding number $\mathcal{N}$~\cite{EdgeDirac2011,ShortTopological,proof_1D}. This is called the bulk-boundary correspondence~\cite{EdgeDirac2011,defect_Jeffrey2010,topo_class}, and hints that the chosen gauge is physically sufficient in characterizing the topological zero modes. 

\begin{figure}
\centering
\includegraphics[width=0.42\textwidth]{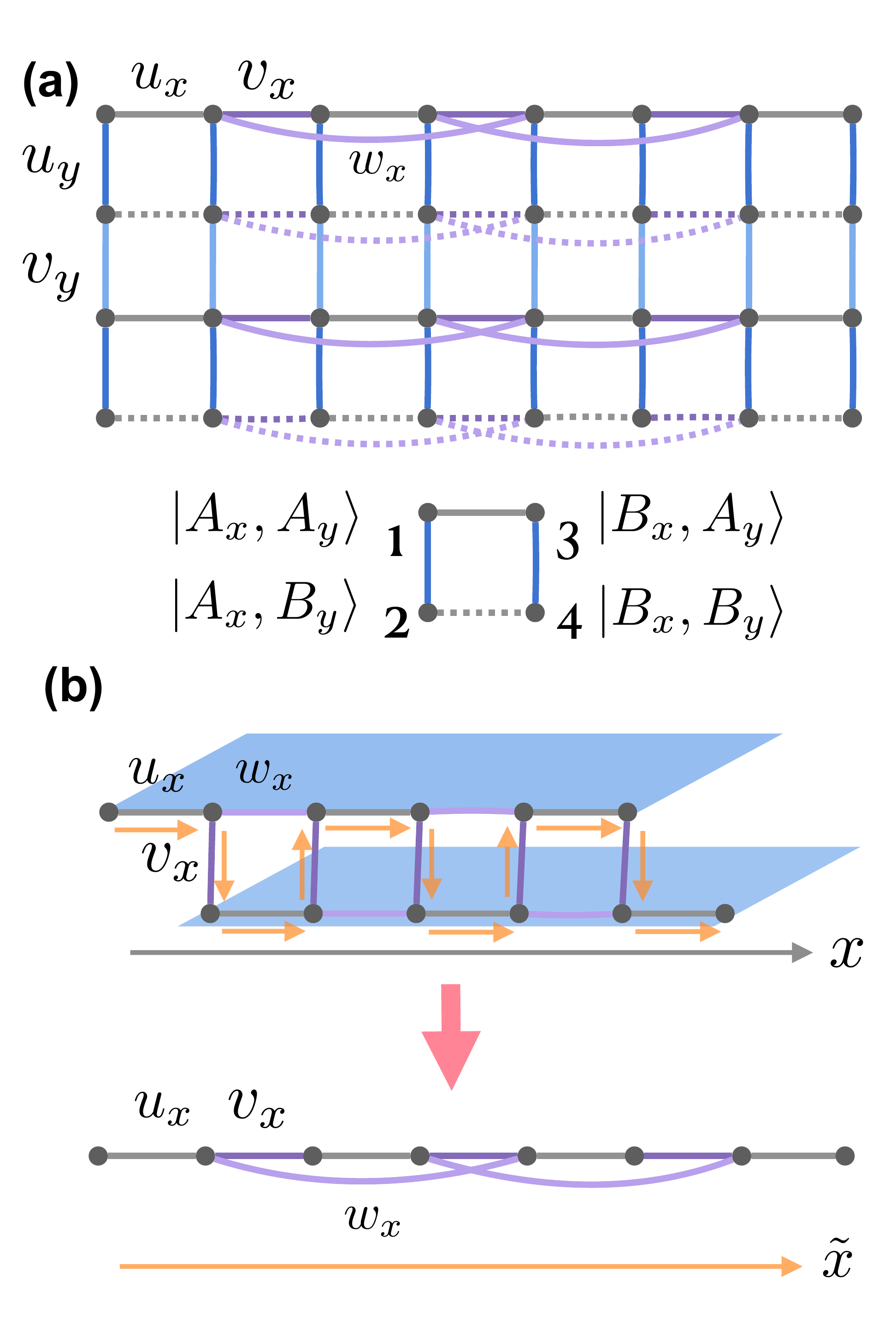}
\caption{(a) A generalized BBH model with an additional long-range hopping in the $x$-direction. The dashed lines correspond to the negative hopping terms with minus signs. (b) Coupling two BBH lattice layers with a stacking fault or AB stacking pattern.}\label{fig:2D_diagram1}
\end{figure}

\section{The Generalized BBH Model}
\label{sec:general}
In Refs.~\cite{multipole_PRB_2017,multipole_science2017}, the authors propose a 2D crystal with vanishing bulk dipole moments and a non-zero quantized off-diagonal quadrupole moment. That model can be regarded as a network composed of perpendicular SSH chains arranged in a square lattice with a $\pi$ flux threading each square plaquette. The resulting spectrum consists of four bands, and the system is insulating at half-filling whenever the hopping parameters in the $x$ and $y$ directions are different. In this paper, we generalize this model by replacing the SSH chains with 1D two-band insulators described in Sec.~\ref{sec:1D} above. The corresponding Bloch Hamiltonian is
\begin{align}
\mathcal{H}(k_x,k_y)=h_x(k_x)\otimes\tau_3 +\sigma_0\otimes h_y(k_y),\label{hBBH}
\end{align}
where $h_x(k_x)=\vec{b}_x(k_x)\cdot\vec{\sigma}$, $h_y(k_y)=\vec{b}_y(k_y)\cdot\vec{\tau}$. The four sub-lattice sites in one unit cell are denoted as $|\alpha_x,\beta_y\rangle$  ($\alpha,\beta=A \text{ or } B$) and shown in Fig.~\ref{fig:2D_diagram1}(a). The Pauli matrix $\tau_3$ in the Hamiltonian $\mathcal{H}(\vec{k})$ reflects the presence of a $\pi$ flux. The characteristic vectors for each of the $h_\mu$ Hamiltonians ($\mu=x,y$)  can be written as 
\begin{align}
\vec{b}_\mu(k_\mu)&=|\vec{b}_\mu(k_\mu)|\big(\cos{\phi_\mu(k_\mu)},\sin{\phi_\mu(k_\mu)},0\big),
\end{align} 
and the eigen-energies of $\mathcal{H}(\vec{k})$ are 
\begin{equation}
 E_\pm(\vec{k})=\pm\sqrt{|\vec{b}_x(k_x)|^2+|\vec{b}_y(k_y)|^2}.\label{E2D}
\end{equation} 
The existence of the two bands at opposite energies is caused by an anticommuting chiral symmetry operation, {\it i.e.} the symmetry $\{\mathcal{H}(\vec{k}),\sigma_3\otimes\tau_3\}=0$. The special separable form of the model of Eq.~(\ref{hBBH}) also has a symmetry as $\{\mathcal{H}(\vec{k}),\sigma_1\otimes\tau_2\mathcal{K}\}=0$ (where $\mathcal{K}$ denotes complex conjugation). Applying these two symmetries, we can prove that both bands are twofold degenerate due to a sub-lattice symmetry,  {\it i.e.} $[\mathcal{H}(\vec{k}),\sigma_2\otimes\tau_1\mathcal{K}]=0$. Note that all these (anti-)symmetries do not connect the Hamiltonian at different values of $\vec{k}$ points and cannot be represented as a spatial symmetry. The eigenstates of the two degenerate negative-energy bands can be expressed as tensor products of the form
\begin{align}
& \Psi_{1}(\vec{k})=\psi^x_{+}(k_x)\otimes\varphi^{1}_-(k_x,k_y),
\nonumber\\
& \Psi_{2}(\vec{k})=\psi^x_{-}(k_x)\otimes\varphi^{2}_-(k_x,k_y)
\,.\label{Psi12}
\end{align}
Here, $\psi^x_{\pm}(k_x)$ are the eigenstates of $h_x(k_x)$ with energies $\pm |\vec{b}_x(\vec{k})|$, respectively. Furthermore, $\varphi^{n}_-(\vec{k})$ ($n=1,2$) is the eigenstate with the negative energy of the $2\times2$ Hamiltonian
\begin{equation}
    h_{xy}^n(\vec{k})=\vec{b}^n_{xy}(\vec{k})\cdot\vec{\tau}
\end{equation} whose characteristic vector is
\begin{equation}
    \vec{b}^n_{xy}(\vec{k})=\left(b^1_{y}(k_y),b^2_{y}(k_y),(-1)^{n-1}|\vec{b}_{x}(k_x)|\right).
\end{equation} 
For later discussion, we also define the polar angle of the characteristic vectors above as 
\begin{align}
\theta_{xy}(\vec{k})=\arccos\left(|\vec{b}_x|/\sqrt{\vec{b}_{x}^2+\vec{b}_{y}^2}\right)\,.
\end{align}
In passing, we note that the generalized BBH model described by Eq.~(\ref{hBBH}) is a special case of the “separability-preserved” Hamiltonians proposed in Ref.~\cite{benalcazar2022chiral}. 

For a 2D crystal, the bulk dipole moments in the directions $\mu=x,y$ are determined by the 2D Zak phases~\cite{RMP_Resta1994,multipole_science2017,multipole_PRB_2017,2DSSH2017} as
\begin{align} \label{eq:polarization_mu_definition}
p_\mu&=\frac{1}{2\pi}\int_{\text{BZ}}\,\Tr[\mathcal{A}_{\mu}(\vec{k})]\,\dd^2\vec{k}\,\text{mod}\,1,
\end{align}
In the SM~\cite{SM}, we show that, for the generalized BBH model, the dipole moments  are equal to
\begin{align}
p_\mu&=\mathcal{N}_\mu\,\text{mod}\,1,
\end{align}
which are always trivial.
The integrals~(\ref{eq:polarization_mu_definition}) of the Berry connections $\mathcal{A}_\mu$ over the occupied bands are the 2D Zak Berry phases, and $\mathcal{N}_\mu$  are the winding numbers of the loops $k_\mu\to\vec{b}_\mu(k_\mu),k_\mu\in[0,2\pi]$.

An important part of our work is answering how to characterize the quadrupole moment of the generalized BBH model. To do that, one first has to obtain the Wilson loops~\cite{multipole_PRB_2017,multipole_science2017}. Using the definitions given above, we obtain an exact result for the Wilson loops along the $x$ and $y$ directions starting from the base point
$\vec{k}=(k_x,k_y)$,
\begin{align}
  \mathcal{W}_{x,\vec{k}}(k_y)&= \exp\left[i \int_{0}^{2\pi}d\phi_x(k'_x)\frac{s_0+\sin{\theta_{xy}(k'_x,k_y)}s_1}{2}\right],\\
  \mathcal{W}_{y,\vec{k}}(k_x)&=\exp\left[i \int_{0}^{2\pi}d\phi_y(k'_y)\frac{s_0-\cos{\theta_{xy}(k_x,k'_y)s_3}}{2}\right].
  \end{align}
 Here we define an additional set of Pauli matrices  $\vec{s}=(s_1,s_2,s_3)$ where the two spinor components correspond to $\Psi_1(\vec{k})$, $\Psi_2(\vec{k})$ respectively and $s_0$ is a $2\times2$ identity matrix. We give the details of the derivation in the SM~\cite{SM}. The above expression allows us to solve the eigen equation 
 \begin{align}
   \mathcal{W}_{\mu,\vec{k}}(k_{\bar{\mu}})\,v_j^\mu(\vec{k})=e^{i2\pi \nu^j_\mu(k_{\bar{\mu}})}\,v_j^\mu(\vec{k}),\label{wilson_loop}
 \end{align}
where $\bar{\mu}=y$ (${\bar \mu} = x$) when $\mu=x$ ($\mu = y$), and  $j=\pm$ labels two eigen-solutions of this equation. The exponents of the eigenvalues above are
\begin{align}
   \nu^j_\mu(k_{\bar{\mu}})&=\int_0^{2\pi}\frac{dk_\mu}{{4\pi}}\frac{d\phi_\mu}{dk_\mu}\left[1+j\frac{|\vec{b}_{\bar{\mu}}(k_{\bar{\mu}})|}{\sqrt{\vec{b}_x^2(k_x)+\vec{b}^2_y(k_y)}}\right]\label{nu_u},
  \end{align}
which are called the Wannier centers. The eigenstates $v_j^\mu(\vec{k})$ with the corresponding Wannier center $\nu_\mu^j(k_{\bar{\mu}})$ generate the Wannier bands of the Wilson loop $\mathcal{W}_{\mu,\vec{k}}(k_{\bar{\mu}})$, not to be confused with the Wannier functions defined later in Sec.~\ref{sec:qinvarant}.

\begin{figure*}
\centering
\includegraphics[width=\textwidth]{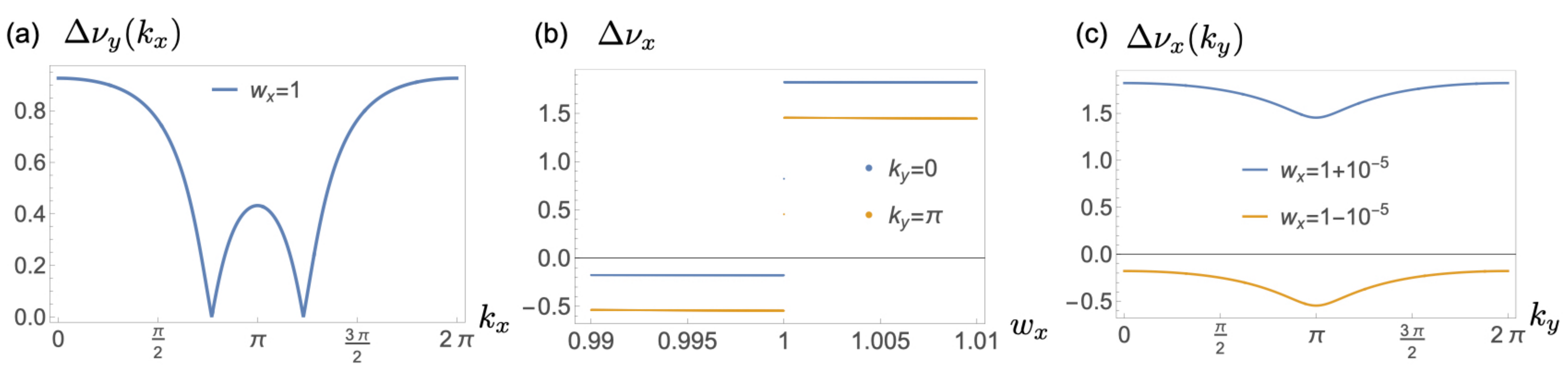}
\caption{Analysis of the Wannier gaps $\Delta\nu_x$ and $\Delta\nu_y$ for the model described by Eq.~(\ref{model_example}), with the nearest-neighbor hoppings as shown in Fig.~\ref{fig:2D_diagram1} chosen as $u_x=u_y=1$ and $v_x=v_y=1.5$. (a) The Wannier gap $\Delta\nu_y$ closes at $w_x=1$, and the winding number $\mathcal{N}_x$ changes from $0$ to $2$.   (b): Discontinuous deformation of the Wannier gap $\Delta\nu_x$ at $k_y=0$ and $k_y=\pi$ when the parameters $w_x$ increases from 0.99 to 1.01. (c): The Wannier gap $\Delta\nu_x(k_y)$ below and above the transition point at $w_x=1$. The yellow and blue lines correspond to $w_x=1 - 10^{-5}$ and $w_x=1 + 10^{-5}$ respectively.}\label{fig:transition}
\end{figure*}

There are two Wannier gaps defined as $\Delta\nu_\mu(k_{\bar{\mu}})=\nu_\mu^+(k_{\bar{\mu}})-\nu_\mu^-(k_{\bar{\mu}})$ for the Wilson loops $\mathcal{W}_{\mu\vec{k}}(k_{\bar{\mu}})$ in the $\mu=x,y$ direction. The gap $\Delta\nu_\mu(k_{\bar{\mu}})$ closes when the winding number $\mathcal{N}_{\bar{\mu}}$ for the perpendicular direction changes. This is due to the fact that $\Delta\nu_\mu(k_{\bar{\mu}})\propto|\vec{b}_{\bar{\mu}}(k_{\bar{\mu}})|$ (see Eq. (\ref{nu_u})) and the closed loop $k_{\bar{\mu}}\in[0,2\pi]\to\vec{b}_{\bar{\mu}}(k_{\bar{\mu}})$ moves through the origin when its winding number changes. This change of winding number $\mathcal{N}_{\bar{\mu}}$ also causes a discontinuous deformation of the other Wannier gap $\Delta\nu_{\bar \mu}(k_\mu)$, even though  in general it does not close, as illustrated in Fig.~\ref{fig:transition}; see the detailed discussion below. The reason for this discontinuous deformation of the Wannier gap $\Delta\nu_{\bar \mu}(k_\mu)$ is that the range of polar angle function $\phi_{\bar{\mu}}(k_{\bar{\mu}})$ in the integral in Eq.~(\ref{nu_u}) suddenly changes, reflecting  the change in winding number.  In conclusion, a transition in ${\cal N}_\mu$ is associated with a discontinuity in $\Delta\nu_\mu$, {\it i.e.}, the Wannier gap in the same direction. At the same time, the Wannier gap $\Delta\nu_{\bar \mu}$ in the perpendicular direction closes. Since the transition from one topological class to another must cause a discontinuity in the Wannier bands, their topology can be described by the set of winding number $(\mathcal{N}_x,\mathcal{N}_y)$. These are defined in terms of two independent 2D Zak phases. Thus, the Wannier topology can be classified as belonging to a $\mathbb{Z}\times \mathbb{Z}$ set. 

In each direction, the Wannier bands are non-degenerate. This result corresponds to the classical case when two dipoles are oriented in the same direction $\mu$ and are separated in the orthogonal direction $\bar{\mu}$, but the total polarization in a unit cell is trivial ({\it i.e.}, it is an integer). Hence it is possible to define the quadrupole moment as in the standard BBH model through the polarizations within a sector of the Wannier bands~\cite{multipole_PRB_2017,multipole_science2017}. This polarization can be related to the 2D Zak-Berry phases of the Wannier basis, which is  defined as
\begin{align}
w^\mu_j(\vec{k})=\big(\Psi_1(\vec{k}), \Psi_2(\vec{k})\big)\cdot v_j^\mu(\vec{k}),
\end{align}
where $\mu$, $j$ and $v_j^\mu(\vec{k})$ are defined  after Eq.~(\ref{wilson_loop}) and in Eq.~(\ref{nu_u}). The polarizations can now be calculated from the formula given in Eq.~(\ref{eq:polarization_mu_definition}), with the trace of the Berry connection $\textrm{Tr}[\mathcal{A}_{\mu}(\vec{k})]$ replaced by $iw^{\bar \mu}_j{}^\dagger(\vec{k}) \partial_{k_\mu} w^{\bar \mu}_j(\vec{k})$. 
For the generalized BBH model, we can choose $w^x_\pm(\vec{k})=\big(\Psi_1(\vec{k}) \pm \Psi_2(\vec{k})\big)/\sqrt{2}$ and $w^y_{\pm}(\vec{k})=\Psi_{1,2}(\vec{k})$ (see details in the SM~\cite{SM}). With these Wannier bases, we find that the Wannier-sector polarizations are linked to the winding numbers of the 1D chains in the $x$- and $y$-directions:
\begin{align}
p_{\mu}^{j}&=\frac{\mathcal{N}_\mu}{2}\,\text{mod}\,1.\label{sec_p}
\end{align}

We then obtain an exact expression for the off-diagonal quadrupole moment,
\begin{align}
q_{xy}=\sum_{j=\pm}p_x^{j}p_y^{j}=\frac{\mathcal{N}_x\mathcal{N}_y}{2}\,\text{mod}\,1.\label{qxy_eq}
\end{align}
The above results, especially Eq.~(\ref{sec_p}), clearly show that the Wannier-sector dipole moments $(p^{\pm}_x,p^{\pm}_y)$ can take the values in a $\mathbb{Z} _2\times\mathbb{Z} _2$ set. In turn, the quadrupole moment is quantized as $q_{xy}=0$ or $1/2$.  This conclusion is consistent with the results from Ref.~\cite{2020DisEQI} where the authors use the real-space operator to understand the quadrupole moment and show that additional spatial symmetry is not necessary for the quantization of the quadrupole moment, even in a disordered system.

As an example of the generalized 2D quadrupole model with chiral symmetry beyond the standard BBH model, we introduce a long-range hopping term in the $x$ direction, as shown in Fig.~\ref{fig:2D_diagram1}(a). The corresponding characteristic vectors are
\begin{align}
\vec{b}_x(k_x)=(&u_x+v_x\cos{k_x}+w_x\cos{(2k_x)},\nonumber\\&v_x\sin{k_x}+w_x\sin{(2k_x)},0),\\ \vec{b}_y(k_y)=(&u_y+v_y\cos{k_y},v_y\sin{k_y},0),   \label{model_example}
\end{align}
 where $u_\mu$, $v_\mu$ are the nearest-neighbor hopping terms, and $w_x$ is the long-range hopping along the $x$ direction. This may seem unrealistic since long-range hopping is usually small and can be neglected in real materials. A possible way to realize such a model is to couple two BBH layers in AB stacking (or a so-called ``stacking fault''~\cite{Partial2019}) pattern and regard the two layers as one. We show a schematic diagram for the two-layer stacking model in Fig.~\ref{fig:2D_diagram1}(b). Apart from the chiral symmetry discussed above, this model is reflection symmetric along the $x$ and $y$ axes.

When $w_x=0$, the model simplifies to the standard BBH model of Refs.~\cite{multipole_PRB_2017,multipole_science2017}. A non-trivial quadrupole moment $q_{xy}=1/2$ exists when $|u_x|<|v_x|$ and $|u_y|<|v_y|$. With the formulas we have derived, we can characterize the Wannier-sector polarization by the winding number in the $x$-direction,  $\mathcal{N}_x$. We find that when $|u_x+w_x|<|v_x|$, $\mathcal{N}_x=1$, while for $|u_x+w_x|>|v_x|$, $\mathcal{N}_x=\text{sign}[(w_x+u_x)(w_x-u_x)]+1$~\cite{SM,ExtendedSSH}. The corresponding $\mathbb{Z}_2$-quantized Wannier-sector polarization $p_x^{\pm}$ is $1/2$ if $|u_x+w_x|<|v_x|$ and becomes $0$ when $|u_x+w_x|>|v_x|$. To be more concrete, we study the case $u_x=u_y=1$, $v_x=v_y=3/2$. As we let $w_x$ increase from 0 to 2, the winding number $\mathcal{N}_x$ switches from $1$ to $0$ at the point $w_x=1/2$ and later switches from $0$ to $2$ at the point $w_x=1$. Note that the Wannier-sector polarization remains trivial for $w_x>1/2$, even when $w_x$ exceeds $1$. As stated above, we expect the gap closure for $\Delta\nu_y(k_x)$ and a discontinuous  behavior of $\Delta\nu_x(k_y)$ at the critical point $w_x=1$, where the winding number $\mathcal{N}_x$ changes from 0 ($1/2<w_x<1$) to 2 ($w_x>1$). The behavior of the gap $\Delta\nu_y(k_x)$ at $w_x=1$ is shown in Fig.~\ref{fig:transition} (a). In Fig.~\ref{fig:transition}(b), we show the Wannier gap $\Delta\nu_x$ as a function of $w_x$ at $k_y=0,\pi$  to highlight its discontinuous change when $w_x$ crosses 1. We also show the Wannier gap $\Delta\nu_x$ on both sides of $w_x=1$, for parameters $w_x=1 - 10^{-5}$ and $w_x= 1 + 10^{-5}$ in Fig.~\ref{fig:transition}(c). Clearly, these figures provide substantial evidence that when $\mathcal{N}_{\mu}$ ($\mu=x,y$) changes, the Wannier gap $\Delta\nu_{\mu}(k_{\bar{\mu}})$ ($\bar{\mu}=y,x$) jumps suddenly. These results are consistent with our previous analysis of the Wannier band topology.

\section{Quadrupole Invariant and Bulk-boundary correspondence}\label{sec:qinvarant}

So far, we have discussed the Wannier-band topology of the general quadrupole model by calculating the sector polarizations based on the nested-Wilson loop formalism. In analogy to the dipole moment, which can be calculated by taking the expectation of the position operator in the localized Wannier functions in a central unit cell, one can obtain an alternative expression of the bulk quadrupole by evaluating the expectation value of the square of the position operators using the same functions~\cite{localizedWannier1997,Resta2011,RMPVanderbilt2012,EQM2020,EQM2021,Watanabe2019,Watanabe2020,Quadrupole_Wannier2021}. In Ref.~\cite{Quadrupole_Wannier2021}, the authors have used this definition to calculate the quadrupole moment and the corner charge in the Wannier representation. They have applied a numerical method and obtained the same value of the corner charge for the BBH model as in the original papers~\cite{multipole_PRB_2017,multipole_science2017}. They also show the quadrupole moment to be gauge-dependent. Recalling that the definition of the quadrupole moment in the Wilson-loop formalism is gauge-invariant, we want to know in what gauge the quadrupole in the Wannier representation is equal to that of the Wilson formalism. We believe this cannot be answered in the general case, but we can solve this issue analytically with the convenient separable form Eq.~(\ref{hBBH}) of the generalized BBH model.

\begin{figure}
\centering
\includegraphics[width=\columnwidth]{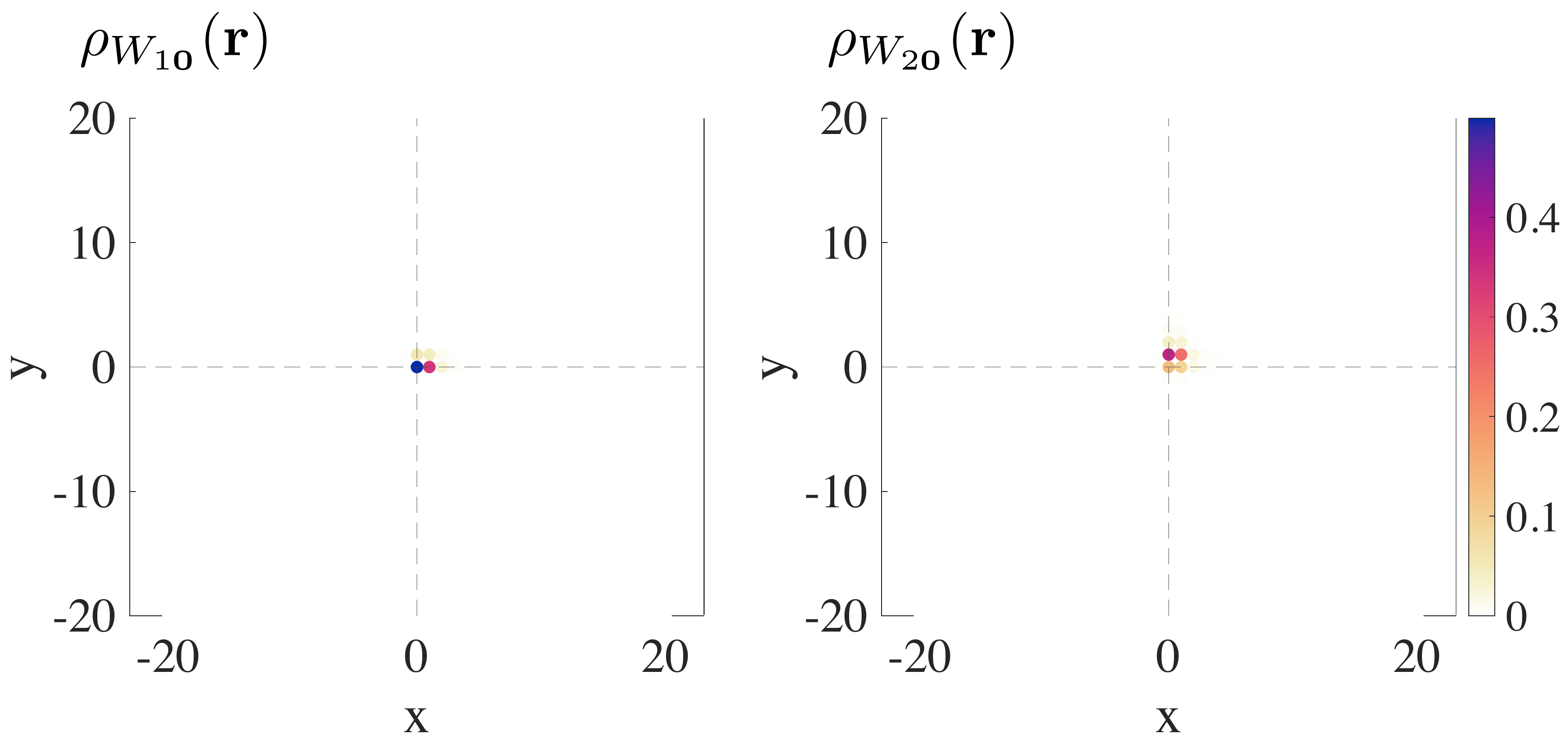}
\caption{Probability density distributions of the Wannier functions $W_{1\vec{0}}$ and $W_{2\vec{0}}$ of the central unit cell over two degenerate bands. The left panel is for $W_{1\vec{0}}$ and the right panel is for $W_{2\vec{0}}$. The parameters in use are $u_x=u_y=1$, $v_x=v_y=3/2$, and $w_x=0$.}\label{fig:2D_diagram2}
\end{figure}

By transforming the quadrupole moment as defined in Ref.~\cite{Quadrupole_Wannier2021} from the Wannier representation to the Bloch representation, we can define the bulk quadrupole tensor as follows
\begin{align}
\mathcal{N}_{\mu\nu}&=\int_{\text{all}} \dd^2\vec{r} \,x_{\mu}x_{\nu}\, W_{n\vec{0}}^\ast(\vec{r}) W_{n\vec{0}}(\vec{r})\nonumber\\
&=-\sum_{n}^{N_\text{occ}}\int_{\text{FBZ}} \frac{d^2\vec{k}}{(2\pi)^2}\Psi^\dagger_{n}(\vec{k})\,\frac{\partial^2}{\partial k_\mu\partial k_\nu}\,\Psi_{n}(\vec{k}),\label{Nxy_k}
\end{align}
where the Wannier functions are defined as
\begin{equation}
    W_{n\vec{R}}(\vec{r})=\sum_{k}e^{-i\vec{k}\cdot\vec{R}}\Psi_{n\vec{k}}(\vec{r})/(N_xN_y).
\end{equation}
The derivation of the above equation is given in the SM~\cite{SM}. The benefit of Eq.~(\ref{Nxy_k}) is that it allows a physical interpretation of the nature of the quadrupole moment since this form is consistent with the quadrupole tensor definition in classical electrostatics. We refer to the Wannier functions $W_{1\vec{0}}(\vec{r})$ and $W_{2\vec{0}}(\vec{r})$ at the central unit cell where $\vec{R}=\vec{0}$ as the central Wannier functions over the bands $n=1,2$. These standard Wannier functions are not related to the Wannier bases we have constructed in Sec.~\ref{sec:general}. We show the density distributions of the Wannier functions $W_{1\vec{0}}(\vec{r})$ and $W_{2\vec{0}}(\vec{r})$ for the standard BBH model with parameters $u_x=u_y=1$, $v_x=v_y=3/2$ and $w_x=0$ in Fig.~\ref{fig:2D_diagram2}. We clearly see that these functions are localized as required.

Applying the gauge of the eigenstates in Eq.~(\ref{Nxy_k}) to the generalized BBH model, we obtain an exact quantized result for the off-diagonal element
\begin{align}
\mathcal{N}_{xy}&=\frac{\mathcal{N}_x\mathcal{N}_y}{2}.\label{eq:Nxy}
\end{align}
This number is unchanged under any continuous deformation of parameters in Eq.~(\ref{hBBH}) that does not change the Wannier topology. Let us call this the quadrupole invariant to distinguish from the quadrupole moment $q_{xy}$. The fractional part of the result \eqref{eq:Nxy} is equal to the quadrupole moment calculated through the Wannier-sector polarization of the generalized BBH model. We can then relate $\mathcal{N}_{xy}$ to $q_{xy}$ by 
\begin{align}
q_{xy}=\mathcal{N}_{xy}\,\text{mod}\,1.
\end{align}
This relation builds the connection between the two definitions of the bulk quadrupole in Ref.~\cite{multipole_PRB_2017} and Ref.~\cite{Quadrupole_Wannier2021}.  We also give expressions for the diagonal elements $\mathcal{N}_{xx}$ and $\mathcal{N}_{yy}$ of the quadrupole tensor in the SM~\cite{SM}. These are not quantized and will change continuously as the system's parameters are tweaked.  

We know that the number of zero modes at every corner of a finite square lattice is $\mathcal{N}_x\mathcal{N}_y$ based on the analysis of the lattice Hamiltonian~\cite{SecondChiral2019,Shin2019,Shin2021}. This can be explained simply  as follows: The Hamiltonian Eq. (\ref{hBBH}) can be realized in real space as the lattice Hamiltonian $H=H_x \otimes I_{N_y} \otimes\tau_z+I_{N_x}\otimes \sigma_0 \otimes H_y$, where $H_\mu$ is the corresponding lattice realization of $h_\mu$. For nontrivial topological phases, all the corner states with zero energy are tensor products of the zero modes of $H_x$ and $H_y$. Thus, the number of zero-modes at one corner is the multiplication of the winding numbers of $H_x$ and $H_y$, that is $\mathcal{N}_c=\mathcal{N}_{x}\mathcal{N}_{y}=2\mathcal{N}_{xy}$. As the existence of the corner solitons shows the loss of the degrees of freedom in the extended states, the corresponding fractional charge at one corner in a half-filling and chiral-symmetry system should be equal to
\begin{align}
Q_c=\frac{\mathcal{N}_c}{2}\,\text{mod}\,1=\mathcal{N}_{xy}\,\text{mod}\,1=q_{xy}.
\end{align}
Thus, we here give a proof of the bulk-boundary correspondence for the generalized BBH model, which holds true beyond the BBH model of Refs.~\cite{multipole_PRB_2017,multipole_science2017} and the numerical discussion of Ref.~\cite{Quadrupole_Wannier2021}.

In Ref.~\cite{benalcazar2022chiral}, the authors have introduced a new bulk invariant called the ``quadrupole chiral number” to characterize the number of second-order topological states of the systems which can have an arbitrary chiral-symmetric Hamiltonian. The link between their  quadrupole chiral number and the physical quadrupole moment in the Wannier and Wilson representations is as yet unclear. Since our result proves the number of corner states is related to the physical quadrupole moment in the generalized BBH model, the connection between the quadrupole chiral number in Ref.~\cite{benalcazar2022chiral}, and the bulk quadrupole moment has  been established within this model.

\section{Gauge issues}
The quadrupole invariant defined in Eq.~(\ref{Nxy_k}) is a gauge-dependent quantity. We fix the gauge by choosing the eigenstates $\Psi_n(\vec{k})$ as Eq.~(\ref{Psi12}). However, a further $\vec{k}$-dependent gauge transformation can break the quantization of this quadrupole invariant, as was stated in Ref.~\cite{Watanabe2019}. The gauge-fixing applied must have a physical basis since it does capture the Wannier topology and boundary properties, as we have shown. We have no analytical insight into this issue at the moment, but the following discussion provides plausible arguments for the validity of our choice.

First of all, the bulk Wannier states of a full periodic system must be chosen to have a consistent gauge with the edge states of a ribbon in the periodic direction \cite{Quadrupole_Wannier2021}. This can be justified by the fact that in a ribbon system, the bulk-like extended states deep in the interior region are not modified in the periodic direction. To make the gauge of bulk and edge states consistent, we require the projection of the bulk central Wannier functions onto the edges to give the same probability density as the central Wannier functions of the edge states in a ribbon that is periodic in the $x$ or $y$ direction. We call this choice the ``edge-consistent gauge''. For the generalized BBH model, we use the gauge in Eq.~ (\ref{1d_gauge}) to describe the 1D edge states in the ribbons and the gauge in Eq.~(\ref{Psi12}) for the 2D bulk states. We have checked that this gauge choice is edge-consistent by numerical calculation, as shown in the SM~\cite{SM}. Since the chosen gauge of the edge states gives the correct number of the corner states discussed in Sec.~\ref{sec:qinvarant}, the consistent gauge of the bulk states should also be able to capture this. 
 
Secondly, in several previous studies of different types of BBH models~\cite{multipole_PRB_2017,multipole_science2017,anomalousHOTs2018,type2_2020,Changan2020}, the fractional charge at the corners $Q_\text{c}$ has been constructed as the summation of the corresponding edge polarizations and the bulk quadrupole moment $Q_\text{c}=p^{\text{edge}}_x+p^{\text{edge}}_y-q_{xy}$. This formula, however, has been given without specifying a common gauge between the Wannier functions of the periodic and ribbon systems, as pointed out in Ref.~\cite{Quadrupole_Wannier2021}. In the SM \cite{SM}, we calculate $Q_\text{c}$ in the edge-consistent gauge we use. We find that
\begin{equation}
Q_\text{c}=p^{\text{edge}}_x=p^{\text{edge}}_y=q_{xy}=\frac{\mathcal{N}_{x}\mathcal{N}_{y}}{2}\,\text{mod}\,1,
\end{equation} 
{\it i.e.} all quantities are half-quantized in the topological phase. This result agrees with the BBH one. In Ref.~\cite{multipole_PRB_2017,multipole_science2017,Changan2020,benalcazar2022chiral}, it is shown that 
in that case $p^{\text{edge}}_x=p^{\text{edge}}_y=Q_c=q_{xy}=1/2$ in the topological phase. 
Thus, it seems that gauge fixing has at least been implicitly assumed there.

Finally, we observe that the gauge of the 1D edge states can only be changed in one direction by a phase factor $U_\mu=\exp{[-i\Theta_\mu(k_\mu)]}$ ($\mu=x,y$). Hence, the corresponding gauge transformation of 2D bulk states must be the multiplication of these transformations, such that $U=U_x(k_x)U_y(k_y)$. Using the gauge-invariant quantum metric tensor, we can prove that the quadrupole invariant remains quantized and that its fractional part is unchanged under the action of factorized gauge transformations such as $U$ \cite{SM}. This seems to imply that the existence of a physical gauge relies on the separability of the model.

\section{Conclusion}
In conclusion, we have provided exact expressions of the Wannier-sector polarizations and quadrupole moment of the generalized BBH model through the Wilson loop formalism and revealed the $\mathbb{Z} \times\mathbb{Z}$ topology of the Wannier bands. This result gives the quantized quadrupole moment without any additional spatial symmetry through BBH's original framework. Furthermore, we have built the connection between two known definitions of the quadrupole moment for our model and made the bulk-boundary correspondence explicit if an edge-consistent gauge is assumed.
 
Quantization of the quadrupole moment in more general situations and how to resolve the gauge redundancy of its alternative definition will be addressed in future studies, though we have provided arguments of plausibility for our choice. Since the quadrupole invariant is related to the quantum metric tensor~\cite{RMP_Resta1994,geometry_metric2000,Resta2011}, which has attracted attention in recent studies~\cite{GMT2018,QuantumMetric2019,EQM2020,EQM2021}, the relation between the higher-order topology and the underlying quantum geometry also requires further investigation. 

The BBH model has been intensely investigated in recent years in various experimental setups~\cite{photonic2019,photonic2020,phononic2018,Acoustic2020,microwave_circuit2018,circuit2018,circuit2019}. Therefore, similar experimental methods could be used to realize the generalized quadrupole model we have discussed, which presents richer second-order topological phases, by coupling two BBH layers in a stacking fault pattern with an inter-layer hopping term.

\acknowledgements
L.Y. acknowledges funding through the China Scholarship Council under Grant No. 201906230305; A.P. acknowledges support from the Leverhulme Trust under the Grant RPG-2019-363. A.P. has received funding from the European Union’s Horizon 2020 research and innovation program under the Marie Sk\l{}odowska-Curie Grant Agreement No 873028. N.R.W. is supported by STFC Grant No. ST/P004423/1. L.Y. thanks Aleksandr Kazantsev, Feng Liu, and Zhiguo Lv for valuable discussions.

\bibliography{BBH.bib}

\begin{thebibliography}{63}%
\makeatletter
\providecommand \@ifxundefined [1]{%
 \@ifx{#1\undefined}
}%
\providecommand \@ifnum [1]{%
 \ifnum #1\expandafter \@firstoftwo
 \else \expandafter \@secondoftwo
 \fi
}%
\providecommand \@ifx [1]{%
 \ifx #1\expandafter \@firstoftwo
 \else \expandafter \@secondoftwo
 \fi
}%
\providecommand \natexlab [1]{#1}%
\providecommand \enquote  [1]{``#1''}%
\providecommand \bibnamefont  [1]{#1}%
\providecommand \bibfnamefont [1]{#1}%
\providecommand \citenamefont [1]{#1}%
\providecommand \href@noop [0]{\@secondoftwo}%
\providecommand \href [0]{\begingroup \@sanitize@url \@href}%
\providecommand \@href[1]{\@@startlink{#1}\@@href}%
\providecommand \@@href[1]{\endgroup#1\@@endlink}%
\providecommand \@sanitize@url [0]{\catcode `\\12\catcode `\$12\catcode
  `\&12\catcode `\#12\catcode `\^12\catcode `\_12\catcode `\%12\relax}%
\providecommand \@@startlink[1]{}%
\providecommand \@@endlink[0]{}%
\providecommand \url  [0]{\begingroup\@sanitize@url \@url }%
\providecommand \@url [1]{\endgroup\@href {#1}{\urlprefix }}%
\providecommand \urlprefix  [0]{URL }%
\providecommand \Eprint [0]{\href }%
\providecommand \doibase [0]{https://doi.org/}%
\providecommand \selectlanguage [0]{\@gobble}%
\providecommand \bibinfo  [0]{\@secondoftwo}%
\providecommand \bibfield  [0]{\@secondoftwo}%
\providecommand \translation [1]{[#1]}%
\providecommand \BibitemOpen [0]{}%
\providecommand \bibitemStop [0]{}%
\providecommand \bibitemNoStop [0]{.\EOS\space}%
\providecommand \EOS [0]{\spacefactor3000\relax}%
\providecommand \BibitemShut  [1]{\csname bibitem#1\endcsname}%
\let\auto@bib@innerbib\@empty
\bibitem [{\citenamefont {Su}\ \emph {et~al.}(1979)\citenamefont {Su},
  \citenamefont {Schrieffer},\ and\ \citenamefont {Heeger}}]{SSH_1979}%
  \BibitemOpen
  \bibfield  {author} {\bibinfo {author} {\bibfnamefont {W.~P.}\ \bibnamefont
  {Su}}, \bibinfo {author} {\bibfnamefont {J.~R.}\ \bibnamefont {Schrieffer}},\
  and\ \bibinfo {author} {\bibfnamefont {A.~J.}\ \bibnamefont {Heeger}},\
  }\bibfield  {title} {\bibinfo {title} {Solitons in polyacetylene},\ }\href
  {https://doi.org/10.1103/PhysRevLett.42.1698} {\bibfield  {journal} {\bibinfo
   {journal} {Phys. Rev. Lett.}\ }\textbf {\bibinfo {volume} {42}},\ \bibinfo
  {pages} {1698} (\bibinfo {year} {1979})}\BibitemShut {NoStop}%
\bibitem [{\citenamefont {Jackiw}\ and\ \citenamefont
  {Rebbi}(1976)}]{Solitons}%
  \BibitemOpen
  \bibfield  {author} {\bibinfo {author} {\bibfnamefont {R.}~\bibnamefont
  {Jackiw}}\ and\ \bibinfo {author} {\bibfnamefont {C.}~\bibnamefont {Rebbi}},\
  }\bibfield  {title} {\bibinfo {title} {Solitons with fermion number 1/2},\
  }\href {https://doi.org/10.1103/PhysRevD.13.3398} {\bibfield  {journal}
  {\bibinfo  {journal} {Phys. Rev. D}\ }\textbf {\bibinfo {volume} {13}},\
  \bibinfo {pages} {3398} (\bibinfo {year} {1976})}\BibitemShut {NoStop}%
\bibitem [{\citenamefont {Chen}\ and\ \citenamefont {Chiou}(2020)}]{proof_1D}%
  \BibitemOpen
  \bibfield  {author} {\bibinfo {author} {\bibfnamefont {B.-H.}\ \bibnamefont
  {Chen}}\ and\ \bibinfo {author} {\bibfnamefont {D.-W.}\ \bibnamefont
  {Chiou}},\ }\bibfield  {title} {\bibinfo {title} {An elementary rigorous
  proof of bulk-boundary correspondence in the generalized su-schrieffer-heeger
  model},\ }\href
  {https://doi.org/https://doi.org/10.1016/j.physleta.2019.126168} {\bibfield
  {journal} {\bibinfo  {journal} {Physics Letters A}\ }\textbf {\bibinfo
  {volume} {384}},\ \bibinfo {pages} {126168} (\bibinfo {year}
  {2020})}\BibitemShut {NoStop}%
\bibitem [{\citenamefont {Haldane}(1988)}]{Haldane1988}%
  \BibitemOpen
  \bibfield  {author} {\bibinfo {author} {\bibfnamefont {F.~D.~M.}\
  \bibnamefont {Haldane}},\ }\bibfield  {title} {\bibinfo {title} {Model for a
  quantum hall effect without landau levels: Condensed-matter realization of
  the "parity anomaly"},\ }\href {https://doi.org/10.1103/PhysRevLett.61.2015}
  {\bibfield  {journal} {\bibinfo  {journal} {Phys. Rev. Lett.}\ }\textbf
  {\bibinfo {volume} {61}},\ \bibinfo {pages} {2015} (\bibinfo {year}
  {1988})}\BibitemShut {NoStop}%
\bibitem [{\citenamefont {Qi}\ \emph {et~al.}(2006{\natexlab{a}})\citenamefont
  {Qi}, \citenamefont {Wu},\ and\ \citenamefont {Zhang}}]{QWZmodel}%
  \BibitemOpen
  \bibfield  {author} {\bibinfo {author} {\bibfnamefont {X.-L.}\ \bibnamefont
  {Qi}}, \bibinfo {author} {\bibfnamefont {Y.-S.}\ \bibnamefont {Wu}},\ and\
  \bibinfo {author} {\bibfnamefont {S.-C.}\ \bibnamefont {Zhang}},\ }\bibfield
  {title} {\bibinfo {title} {Topological quantization of the spin hall effect
  in two-dimensional paramagnetic semiconductors},\ }\href
  {https://doi.org/10.1103/PhysRevB.74.085308} {\bibfield  {journal} {\bibinfo
  {journal} {Phys. Rev. B}\ }\textbf {\bibinfo {volume} {74}},\ \bibinfo
  {pages} {085308} (\bibinfo {year} {2006}{\natexlab{a}})}\BibitemShut
  {NoStop}%
\bibitem [{\citenamefont {Kane}\ and\ \citenamefont
  {Mele}(2005{\natexlab{a}})}]{KaneQHSE2005}%
  \BibitemOpen
  \bibfield  {author} {\bibinfo {author} {\bibfnamefont {C.~L.}\ \bibnamefont
  {Kane}}\ and\ \bibinfo {author} {\bibfnamefont {E.~J.}\ \bibnamefont
  {Mele}},\ }\bibfield  {title} {\bibinfo {title} {Quantum spin hall effect in
  graphene},\ }\href {https://doi.org/10.1103/PhysRevLett.95.226801} {\bibfield
   {journal} {\bibinfo  {journal} {Phys. Rev. Lett.}\ }\textbf {\bibinfo
  {volume} {95}},\ \bibinfo {pages} {226801} (\bibinfo {year}
  {2005}{\natexlab{a}})}\BibitemShut {NoStop}%
\bibitem [{\citenamefont {Qi}\ \emph {et~al.}(2006{\natexlab{b}})\citenamefont
  {Qi}, \citenamefont {Wu},\ and\ \citenamefont
  {Zhang}}]{bulk_edge_XiaoLiang2006}%
  \BibitemOpen
  \bibfield  {author} {\bibinfo {author} {\bibfnamefont {X.-L.}\ \bibnamefont
  {Qi}}, \bibinfo {author} {\bibfnamefont {Y.-S.}\ \bibnamefont {Wu}},\ and\
  \bibinfo {author} {\bibfnamefont {S.-C.}\ \bibnamefont {Zhang}},\ }\bibfield
  {title} {\bibinfo {title} {General theorem relating the bulk topological
  number to edge states in two-dimensional insulators},\ }\href
  {https://doi.org/10.1103/PhysRevB.74.045125} {\bibfield  {journal} {\bibinfo
  {journal} {Phys. Rev. B}\ }\textbf {\bibinfo {volume} {74}},\ \bibinfo
  {pages} {045125} (\bibinfo {year} {2006}{\natexlab{b}})}\BibitemShut
  {NoStop}%
\bibitem [{\citenamefont {Moore}\ and\ \citenamefont
  {Balents}(2007)}]{TRSinvariant2007}%
  \BibitemOpen
  \bibfield  {author} {\bibinfo {author} {\bibfnamefont {J.~E.}\ \bibnamefont
  {Moore}}\ and\ \bibinfo {author} {\bibfnamefont {L.}~\bibnamefont
  {Balents}},\ }\bibfield  {title} {\bibinfo {title} {Topological invariants of
  time-reversal-invariant band structures},\ }\href
  {https://doi.org/10.1103/PhysRevB.75.121306} {\bibfield  {journal} {\bibinfo
  {journal} {Phys. Rev. B}\ }\textbf {\bibinfo {volume} {75}},\ \bibinfo
  {pages} {121306} (\bibinfo {year} {2007})}\BibitemShut {NoStop}%
\bibitem [{\citenamefont {Kane}\ and\ \citenamefont
  {Mele}(2005{\natexlab{b}})}]{Kane_Z2}%
  \BibitemOpen
  \bibfield  {author} {\bibinfo {author} {\bibfnamefont {C.~L.}\ \bibnamefont
  {Kane}}\ and\ \bibinfo {author} {\bibfnamefont {E.~J.}\ \bibnamefont
  {Mele}},\ }\bibfield  {title} {\bibinfo {title} {${Z}_{2}$ topological order
  and the quantum spin hall effect},\ }\href
  {https://doi.org/10.1103/PhysRevLett.95.146802} {\bibfield  {journal}
  {\bibinfo  {journal} {Phys. Rev. Lett.}\ }\textbf {\bibinfo {volume} {95}},\
  \bibinfo {pages} {146802} (\bibinfo {year} {2005}{\natexlab{b}})}\BibitemShut
  {NoStop}%
\bibitem [{\citenamefont {Bernevig}\ and\ \citenamefont
  {Zhang}(2006)}]{ShouchengQSHE}%
  \BibitemOpen
  \bibfield  {author} {\bibinfo {author} {\bibfnamefont {B.~A.}\ \bibnamefont
  {Bernevig}}\ and\ \bibinfo {author} {\bibfnamefont {S.-C.}\ \bibnamefont
  {Zhang}},\ }\bibfield  {title} {\bibinfo {title} {Quantum spin hall effect},\
  }\href {https://doi.org/10.1103/PhysRevLett.96.106802} {\bibfield  {journal}
  {\bibinfo  {journal} {Phys. Rev. Lett.}\ }\textbf {\bibinfo {volume} {96}},\
  \bibinfo {pages} {106802} (\bibinfo {year} {2006})}\BibitemShut {NoStop}%
\bibitem [{\citenamefont {Fu}\ \emph {et~al.}(2007)\citenamefont {Fu},
  \citenamefont {Kane},\ and\ \citenamefont {Mele}}]{TI3D_liang2007}%
  \BibitemOpen
  \bibfield  {author} {\bibinfo {author} {\bibfnamefont {L.}~\bibnamefont
  {Fu}}, \bibinfo {author} {\bibfnamefont {C.~L.}\ \bibnamefont {Kane}},\ and\
  \bibinfo {author} {\bibfnamefont {E.~J.}\ \bibnamefont {Mele}},\ }\bibfield
  {title} {\bibinfo {title} {Topological insulators in three dimensions},\
  }\href {https://doi.org/10.1103/PhysRevLett.98.106803} {\bibfield  {journal}
  {\bibinfo  {journal} {Phys. Rev. Lett.}\ }\textbf {\bibinfo {volume} {98}},\
  \bibinfo {pages} {106803} (\bibinfo {year} {2007})}\BibitemShut {NoStop}%
\bibitem [{\citenamefont {Fu}\ and\ \citenamefont
  {Kane}(2007)}]{TIindices_liang2007}%
  \BibitemOpen
  \bibfield  {author} {\bibinfo {author} {\bibfnamefont {L.}~\bibnamefont
  {Fu}}\ and\ \bibinfo {author} {\bibfnamefont {C.~L.}\ \bibnamefont {Kane}},\
  }\bibfield  {title} {\bibinfo {title} {Topological insulators with inversion
  symmetry},\ }\href {https://doi.org/10.1103/PhysRevB.76.045302} {\bibfield
  {journal} {\bibinfo  {journal} {Phys. Rev. B}\ }\textbf {\bibinfo {volume}
  {76}},\ \bibinfo {pages} {045302} (\bibinfo {year} {2007})}\BibitemShut
  {NoStop}%
\bibitem [{\citenamefont {Roy}(2009)}]{QSHE3D_Roy2009}%
  \BibitemOpen
  \bibfield  {author} {\bibinfo {author} {\bibfnamefont {R.}~\bibnamefont
  {Roy}},\ }\bibfield  {title} {\bibinfo {title} {Topological phases and the
  quantum spin hall effect in three dimensions},\ }\href
  {https://doi.org/10.1103/PhysRevB.79.195322} {\bibfield  {journal} {\bibinfo
  {journal} {Phys. Rev. B}\ }\textbf {\bibinfo {volume} {79}},\ \bibinfo
  {pages} {195322} (\bibinfo {year} {2009})}\BibitemShut {NoStop}%
\bibitem [{\citenamefont {Teo}\ and\ \citenamefont
  {Kane}(2010)}]{defect_Jeffrey2010}%
  \BibitemOpen
  \bibfield  {author} {\bibinfo {author} {\bibfnamefont {J.~C.~Y.}\
  \bibnamefont {Teo}}\ and\ \bibinfo {author} {\bibfnamefont {C.~L.}\
  \bibnamefont {Kane}},\ }\bibfield  {title} {\bibinfo {title} {Topological
  defects and gapless modes in insulators and superconductors},\ }\href
  {https://doi.org/10.1103/PhysRevB.82.115120} {\bibfield  {journal} {\bibinfo
  {journal} {Phys. Rev. B}\ }\textbf {\bibinfo {volume} {82}},\ \bibinfo
  {pages} {115120} (\bibinfo {year} {2010})}\BibitemShut {NoStop}%
\bibitem [{\citenamefont {Hasan}\ and\ \citenamefont
  {Kane}(2010)}]{HasanRMP2010}%
  \BibitemOpen
  \bibfield  {author} {\bibinfo {author} {\bibfnamefont {M.~Z.}\ \bibnamefont
  {Hasan}}\ and\ \bibinfo {author} {\bibfnamefont {C.~L.}\ \bibnamefont
  {Kane}},\ }\bibfield  {title} {\bibinfo {title} {Colloquium: Topological
  insulators},\ }\href {https://doi.org/10.1103/RevModPhys.82.3045} {\bibfield
  {journal} {\bibinfo  {journal} {Rev. Mod. Phys.}\ }\textbf {\bibinfo {volume}
  {82}},\ \bibinfo {pages} {3045} (\bibinfo {year} {2010})}\BibitemShut
  {NoStop}%
\bibitem [{\citenamefont {Mong}\ and\ \citenamefont
  {Shivamoggi}(2011)}]{EdgeDirac2011}%
  \BibitemOpen
  \bibfield  {author} {\bibinfo {author} {\bibfnamefont {R.~S.~K.}\
  \bibnamefont {Mong}}\ and\ \bibinfo {author} {\bibfnamefont {V.}~\bibnamefont
  {Shivamoggi}},\ }\bibfield  {title} {\bibinfo {title} {Edge states and the
  bulk-boundary correspondence in dirac hamiltonians},\ }\href
  {https://doi.org/10.1103/PhysRevB.83.125109} {\bibfield  {journal} {\bibinfo
  {journal} {Phys. Rev. B}\ }\textbf {\bibinfo {volume} {83}},\ \bibinfo
  {pages} {125109} (\bibinfo {year} {2011})}\BibitemShut {NoStop}%
\bibitem [{\citenamefont {Chiu}\ \emph {et~al.}(2016)\citenamefont {Chiu},
  \citenamefont {Teo}, \citenamefont {Schnyder},\ and\ \citenamefont
  {Ryu}}]{topo_class}%
  \BibitemOpen
  \bibfield  {author} {\bibinfo {author} {\bibfnamefont {C.-K.}\ \bibnamefont
  {Chiu}}, \bibinfo {author} {\bibfnamefont {J.~C.~Y.}\ \bibnamefont {Teo}},
  \bibinfo {author} {\bibfnamefont {A.~P.}\ \bibnamefont {Schnyder}},\ and\
  \bibinfo {author} {\bibfnamefont {S.}~\bibnamefont {Ryu}},\ }\bibfield
  {title} {\bibinfo {title} {Classification of topological quantum matter with
  symmetries},\ }\href {https://doi.org/10.1103/RevModPhys.88.035005}
  {\bibfield  {journal} {\bibinfo  {journal} {Rev. Mod. Phys.}\ }\textbf
  {\bibinfo {volume} {88}},\ \bibinfo {pages} {035005} (\bibinfo {year}
  {2016})}\BibitemShut {NoStop}%
\bibitem [{\citenamefont {Duncan}\ \emph {et~al.}(2018)\citenamefont {Duncan},
  \citenamefont {\"Ohberg},\ and\ \citenamefont {Valiente}}]{Exact_edge2018}%
  \BibitemOpen
  \bibfield  {author} {\bibinfo {author} {\bibfnamefont {C.~W.}\ \bibnamefont
  {Duncan}}, \bibinfo {author} {\bibfnamefont {P.}~\bibnamefont {\"Ohberg}},\
  and\ \bibinfo {author} {\bibfnamefont {M.}~\bibnamefont {Valiente}},\
  }\bibfield  {title} {\bibinfo {title} {Exact edge, bulk, and bound states of
  finite topological systems},\ }\href
  {https://doi.org/10.1103/PhysRevB.97.195439} {\bibfield  {journal} {\bibinfo
  {journal} {Phys. Rev. B}\ }\textbf {\bibinfo {volume} {97}},\ \bibinfo
  {pages} {195439} (\bibinfo {year} {2018})}\BibitemShut {NoStop}%
\bibitem [{\citenamefont {Benalcazar}\ \emph
  {et~al.}(2017{\natexlab{a}})\citenamefont {Benalcazar}, \citenamefont
  {Bernevig},\ and\ \citenamefont {Hughes}}]{multipole_PRB_2017}%
  \BibitemOpen
  \bibfield  {author} {\bibinfo {author} {\bibfnamefont {W.~A.}\ \bibnamefont
  {Benalcazar}}, \bibinfo {author} {\bibfnamefont {B.~A.}\ \bibnamefont
  {Bernevig}},\ and\ \bibinfo {author} {\bibfnamefont {T.~L.}\ \bibnamefont
  {Hughes}},\ }\bibfield  {title} {\bibinfo {title} {Electric multipole
  moments, topological multipole moment pumping, and chiral hinge states in
  crystalline insulators},\ }\href {https://doi.org/10.1103/PhysRevB.96.245115}
  {\bibfield  {journal} {\bibinfo  {journal} {Phys. Rev. B}\ }\textbf {\bibinfo
  {volume} {96}},\ \bibinfo {pages} {245115} (\bibinfo {year}
  {2017}{\natexlab{a}})}\BibitemShut {NoStop}%
\bibitem [{\citenamefont {Benalcazar}\ \emph
  {et~al.}(2017{\natexlab{b}})\citenamefont {Benalcazar}, \citenamefont
  {Bernevig},\ and\ \citenamefont {Hughes}}]{multipole_science2017}%
  \BibitemOpen
  \bibfield  {author} {\bibinfo {author} {\bibfnamefont {W.~A.}\ \bibnamefont
  {Benalcazar}}, \bibinfo {author} {\bibfnamefont {B.~A.}\ \bibnamefont
  {Bernevig}},\ and\ \bibinfo {author} {\bibfnamefont {T.~L.}\ \bibnamefont
  {Hughes}},\ }\bibfield  {title} {\bibinfo {title} {Quantized electric
  multipole insulators},\ }\href {https://doi.org/10.1126/science.aah6442}
  {\bibfield  {journal} {\bibinfo  {journal} {Science}\ }\textbf {\bibinfo
  {volume} {357}},\ \bibinfo {pages} {61–66} (\bibinfo {year}
  {2017}{\natexlab{b}})}\BibitemShut {NoStop}%
\bibitem [{\citenamefont {Song}\ \emph {et~al.}(2017)\citenamefont {Song},
  \citenamefont {Fang},\ and\ \citenamefont {Fang}}]{Zhida2017}%
  \BibitemOpen
  \bibfield  {author} {\bibinfo {author} {\bibfnamefont {Z.}~\bibnamefont
  {Song}}, \bibinfo {author} {\bibfnamefont {Z.}~\bibnamefont {Fang}},\ and\
  \bibinfo {author} {\bibfnamefont {C.}~\bibnamefont {Fang}},\ }\bibfield
  {title} {\bibinfo {title} {(d-2)-dimensional edge states of rotation symmetry
  protected topological states},\ }\href
  {https://doi.org/10.1103/PhysRevLett.119.246402} {\bibfield  {journal}
  {\bibinfo  {journal} {Phys. Rev. Lett.}\ }\textbf {\bibinfo {volume} {119}},\
  \bibinfo {pages} {246402} (\bibinfo {year} {2017})}\BibitemShut {NoStop}%
\bibitem [{\citenamefont {Schindler}\ \emph {et~al.}(2018)\citenamefont
  {Schindler}, \citenamefont {Cook}, \citenamefont {Vergniory}, \citenamefont
  {Wang}, \citenamefont {Parkin}, \citenamefont {Bernevig},\ and\ \citenamefont
  {Neupert}}]{HOTI_science2018}%
  \BibitemOpen
  \bibfield  {author} {\bibinfo {author} {\bibfnamefont {F.}~\bibnamefont
  {Schindler}}, \bibinfo {author} {\bibfnamefont {A.~M.}\ \bibnamefont {Cook}},
  \bibinfo {author} {\bibfnamefont {M.~G.}\ \bibnamefont {Vergniory}}, \bibinfo
  {author} {\bibfnamefont {Z.}~\bibnamefont {Wang}}, \bibinfo {author}
  {\bibfnamefont {S.~S.~P.}\ \bibnamefont {Parkin}}, \bibinfo {author}
  {\bibfnamefont {B.~A.}\ \bibnamefont {Bernevig}},\ and\ \bibinfo {author}
  {\bibfnamefont {T.}~\bibnamefont {Neupert}},\ }\bibfield  {title} {\bibinfo
  {title} {Higher-order topological insulators},\ }\bibfield  {journal}
  {\bibinfo  {journal} {Science Advances}\ }\textbf {\bibinfo {volume} {4}},\
  \href {https://doi.org/10.1126/sciadv.aat0346} {10.1126/sciadv.aat0346}
  (\bibinfo {year} {2018})\BibitemShut {NoStop}%
\bibitem [{\citenamefont {van Miert}\ and\ \citenamefont
  {Ortix}(2018)}]{HOTinversion2018}%
  \BibitemOpen
  \bibfield  {author} {\bibinfo {author} {\bibfnamefont {G.}~\bibnamefont {van
  Miert}}\ and\ \bibinfo {author} {\bibfnamefont {C.}~\bibnamefont {Ortix}},\
  }\bibfield  {title} {\bibinfo {title} {Higher-order topological insulators
  protected by inversion and rotoinversion symmetries},\ }\href
  {https://doi.org/10.1103/PhysRevB.98.081110} {\bibfield  {journal} {\bibinfo
  {journal} {Phys. Rev. B}\ }\textbf {\bibinfo {volume} {98}},\ \bibinfo
  {pages} {081110} (\bibinfo {year} {2018})}\BibitemShut {NoStop}%
\bibitem [{\citenamefont {Khalaf}(2018)}]{HOTISinversion2018}%
  \BibitemOpen
  \bibfield  {author} {\bibinfo {author} {\bibfnamefont {E.}~\bibnamefont
  {Khalaf}},\ }\bibfield  {title} {\bibinfo {title} {Higher-order topological
  insulators and superconductors protected by inversion symmetry},\ }\href
  {https://doi.org/10.1103/PhysRevB.97.205136} {\bibfield  {journal} {\bibinfo
  {journal} {Phys. Rev. B}\ }\textbf {\bibinfo {volume} {97}},\ \bibinfo
  {pages} {205136} (\bibinfo {year} {2018})}\BibitemShut {NoStop}%
\bibitem [{\citenamefont {Geier}\ \emph {et~al.}(2018)\citenamefont {Geier},
  \citenamefont {Trifunovic}, \citenamefont {Hoskam},\ and\ \citenamefont
  {Brouwer}}]{second_order2}%
  \BibitemOpen
  \bibfield  {author} {\bibinfo {author} {\bibfnamefont {M.}~\bibnamefont
  {Geier}}, \bibinfo {author} {\bibfnamefont {L.}~\bibnamefont {Trifunovic}},
  \bibinfo {author} {\bibfnamefont {M.}~\bibnamefont {Hoskam}},\ and\ \bibinfo
  {author} {\bibfnamefont {P.~W.}\ \bibnamefont {Brouwer}},\ }\bibfield
  {title} {\bibinfo {title} {Second-order topological insulators and
  superconductors with an order-two crystalline symmetry},\ }\href
  {https://doi.org/10.1103/PhysRevB.97.205135} {\bibfield  {journal} {\bibinfo
  {journal} {Phys. Rev. B}\ }\textbf {\bibinfo {volume} {97}},\ \bibinfo
  {pages} {205135} (\bibinfo {year} {2018})}\BibitemShut {NoStop}%
\bibitem [{\citenamefont {C\ifmmode \u{a}\else \u{a}\fi{}lug\ifmmode~\u{a}\else
  \u{a}\fi{}ru}\ \emph {et~al.}(2019)\citenamefont {C\ifmmode \u{a}\else
  \u{a}\fi{}lug\ifmmode~\u{a}\else \u{a}\fi{}ru}, \citenamefont {Juri\ifmmode
  \check{c}\else \v{c}\fi{}i\ifmmode~\acute{c}\else \'{c}\fi{}},\ and\
  \citenamefont {Roy}}]{HOTprinciple2019}%
  \BibitemOpen
  \bibfield  {author} {\bibinfo {author} {\bibfnamefont {D.}~\bibnamefont
  {C\ifmmode \u{a}\else \u{a}\fi{}lug\ifmmode~\u{a}\else \u{a}\fi{}ru}},
  \bibinfo {author} {\bibfnamefont {V.}~\bibnamefont {Juri\ifmmode
  \check{c}\else \v{c}\fi{}i\ifmmode~\acute{c}\else \'{c}\fi{}}},\ and\
  \bibinfo {author} {\bibfnamefont {B.}~\bibnamefont {Roy}},\ }\bibfield
  {title} {\bibinfo {title} {Higher-order topological phases: A general
  principle of construction},\ }\href
  {https://doi.org/10.1103/PhysRevB.99.041301} {\bibfield  {journal} {\bibinfo
  {journal} {Phys. Rev. B}\ }\textbf {\bibinfo {volume} {99}},\ \bibinfo
  {pages} {041301} (\bibinfo {year} {2019})}\BibitemShut {NoStop}%
\bibitem [{\citenamefont {Wieder}\ \emph {et~al.}(2020)\citenamefont {Wieder},
  \citenamefont {Wang}, \citenamefont {Cano}, \citenamefont {Dai},
  \citenamefont {Schoop}, \citenamefont {Bradlyn},\ and\ \citenamefont
  {Bernevig}}]{FermiArcs2020}%
  \BibitemOpen
  \bibfield  {author} {\bibinfo {author} {\bibfnamefont {B.~J.}\ \bibnamefont
  {Wieder}}, \bibinfo {author} {\bibfnamefont {Z.}~\bibnamefont {Wang}},
  \bibinfo {author} {\bibfnamefont {J.}~\bibnamefont {Cano}}, \bibinfo {author}
  {\bibfnamefont {X.}~\bibnamefont {Dai}}, \bibinfo {author} {\bibfnamefont
  {L.~M.}\ \bibnamefont {Schoop}}, \bibinfo {author} {\bibfnamefont
  {B.}~\bibnamefont {Bradlyn}},\ and\ \bibinfo {author} {\bibfnamefont {B.~A.}\
  \bibnamefont {Bernevig}},\ }\bibfield  {title} {\bibinfo {title} {Strong and
  fragile topological dirac semimetals with higher-order fermi arcs},\ }\href
  {https://doi.org/10.1038/s41467-020-14443-5} {\bibfield  {journal} {\bibinfo
  {journal} {Nature Communications}\ }\textbf {\bibinfo {volume} {11}},\
  \bibinfo {pages} {627} (\bibinfo {year} {2020})}\BibitemShut {NoStop}%
\bibitem [{\citenamefont {Ren}\ \emph {et~al.}(2020)\citenamefont {Ren},
  \citenamefont {Qiao},\ and\ \citenamefont {Niu}}]{Yafei2020}%
  \BibitemOpen
  \bibfield  {author} {\bibinfo {author} {\bibfnamefont {Y.}~\bibnamefont
  {Ren}}, \bibinfo {author} {\bibfnamefont {Z.}~\bibnamefont {Qiao}},\ and\
  \bibinfo {author} {\bibfnamefont {Q.}~\bibnamefont {Niu}},\ }\bibfield
  {title} {\bibinfo {title} {Engineering corner states from two-dimensional
  topological insulators},\ }\href
  {https://doi.org/10.1103/PhysRevLett.124.166804} {\bibfield  {journal}
  {\bibinfo  {journal} {Phys. Rev. Lett.}\ }\textbf {\bibinfo {volume} {124}},\
  \bibinfo {pages} {166804} (\bibinfo {year} {2020})}\BibitemShut {NoStop}%
\bibitem [{\citenamefont {Li}\ and\ \citenamefont {Wu}(2020)}]{Changan2020}%
  \BibitemOpen
  \bibfield  {author} {\bibinfo {author} {\bibfnamefont {C.-A.}\ \bibnamefont
  {Li}}\ and\ \bibinfo {author} {\bibfnamefont {S.-S.}\ \bibnamefont {Wu}},\
  }\bibfield  {title} {\bibinfo {title} {Topological states in generalized
  electric quadrupole insulators},\ }\href
  {https://doi.org/10.1103/PhysRevB.101.195309} {\bibfield  {journal} {\bibinfo
   {journal} {Phys. Rev. B}\ }\textbf {\bibinfo {volume} {101}},\ \bibinfo
  {pages} {195309} (\bibinfo {year} {2020})}\BibitemShut {NoStop}%
\bibitem [{\citenamefont {Yang}\ \emph {et~al.}(2020)\citenamefont {Yang},
  \citenamefont {Li}, \citenamefont {Duan},\ and\ \citenamefont
  {Xu}}]{type2_2020}%
  \BibitemOpen
  \bibfield  {author} {\bibinfo {author} {\bibfnamefont {Y.-B.}\ \bibnamefont
  {Yang}}, \bibinfo {author} {\bibfnamefont {K.}~\bibnamefont {Li}}, \bibinfo
  {author} {\bibfnamefont {L.-M.}\ \bibnamefont {Duan}},\ and\ \bibinfo
  {author} {\bibfnamefont {Y.}~\bibnamefont {Xu}},\ }\bibfield  {title}
  {\bibinfo {title} {Type-ii quadrupole topological insulators},\ }\href
  {https://doi.org/10.1103/PhysRevResearch.2.033029} {\bibfield  {journal}
  {\bibinfo  {journal} {Phys. Rev. Research}\ }\textbf {\bibinfo {volume}
  {2}},\ \bibinfo {pages} {033029} (\bibinfo {year} {2020})}\BibitemShut
  {NoStop}%
\bibitem [{\citenamefont {Jung}\ \emph {et~al.}(2021)\citenamefont {Jung},
  \citenamefont {Yu},\ and\ \citenamefont {Shvets}}]{Exact_localized2021}%
  \BibitemOpen
  \bibfield  {author} {\bibinfo {author} {\bibfnamefont {M.}~\bibnamefont
  {Jung}}, \bibinfo {author} {\bibfnamefont {Y.}~\bibnamefont {Yu}},\ and\
  \bibinfo {author} {\bibfnamefont {G.}~\bibnamefont {Shvets}},\ }\bibfield
  {title} {\bibinfo {title} {Exact higher-order bulk-boundary correspondence of
  corner-localized states},\ }\href
  {https://doi.org/10.1103/PhysRevB.104.195437} {\bibfield  {journal} {\bibinfo
   {journal} {Phys. Rev. B}\ }\textbf {\bibinfo {volume} {104}},\ \bibinfo
  {pages} {195437} (\bibinfo {year} {2021})}\BibitemShut {NoStop}%
\bibitem [{\citenamefont {Mittal}\ \emph {et~al.}(2019)\citenamefont {Mittal},
  \citenamefont {Orre}, \citenamefont {Zhu}, \citenamefont {Gorlach},
  \citenamefont {Poddubny},\ and\ \citenamefont {Hafezi}}]{photonic2019}%
  \BibitemOpen
  \bibfield  {author} {\bibinfo {author} {\bibfnamefont {S.}~\bibnamefont
  {Mittal}}, \bibinfo {author} {\bibfnamefont {V.~V.}\ \bibnamefont {Orre}},
  \bibinfo {author} {\bibfnamefont {G.}~\bibnamefont {Zhu}}, \bibinfo {author}
  {\bibfnamefont {M.~A.}\ \bibnamefont {Gorlach}}, \bibinfo {author}
  {\bibfnamefont {A.}~\bibnamefont {Poddubny}},\ and\ \bibinfo {author}
  {\bibfnamefont {M.}~\bibnamefont {Hafezi}},\ }\bibfield  {title} {\bibinfo
  {title} {Photonic quadrupole topological phases},\ }\href
  {https://doi.org/10.1038/s41566-019-0452-0} {\bibfield  {journal} {\bibinfo
  {journal} {Nature Photonics}\ }\textbf {\bibinfo {volume} {13}},\ \bibinfo
  {pages} {692} (\bibinfo {year} {2019})}\BibitemShut {NoStop}%
\bibitem [{\citenamefont {He}\ \emph {et~al.}(2020)\citenamefont {He},
  \citenamefont {Addison}, \citenamefont {Mele},\ and\ \citenamefont
  {Zhen}}]{photonic2020}%
  \BibitemOpen
  \bibfield  {author} {\bibinfo {author} {\bibfnamefont {L.}~\bibnamefont
  {He}}, \bibinfo {author} {\bibfnamefont {Z.}~\bibnamefont {Addison}},
  \bibinfo {author} {\bibfnamefont {E.~J.}\ \bibnamefont {Mele}},\ and\
  \bibinfo {author} {\bibfnamefont {B.}~\bibnamefont {Zhen}},\ }\bibfield
  {title} {\bibinfo {title} {Quadrupole topological photonic crystals},\ }\href
  {https://doi.org/10.1038/s41467-020-16916-z} {\bibfield  {journal} {\bibinfo
  {journal} {Nature Communications}\ }\textbf {\bibinfo {volume} {11}},\
  \bibinfo {pages} {3119} (\bibinfo {year} {2020})}\BibitemShut {NoStop}%
\bibitem [{\citenamefont {Serra-Garcia}\ \emph {et~al.}(2018)\citenamefont
  {Serra-Garcia}, \citenamefont {Peri}, \citenamefont {S{\"u}sstrunk},
  \citenamefont {Bilal}, \citenamefont {Larsen}, \citenamefont {Villanueva},\
  and\ \citenamefont {Huber}}]{phononic2018}%
  \BibitemOpen
  \bibfield  {author} {\bibinfo {author} {\bibfnamefont {M.}~\bibnamefont
  {Serra-Garcia}}, \bibinfo {author} {\bibfnamefont {V.}~\bibnamefont {Peri}},
  \bibinfo {author} {\bibfnamefont {R.}~\bibnamefont {S{\"u}sstrunk}}, \bibinfo
  {author} {\bibfnamefont {O.~R.}\ \bibnamefont {Bilal}}, \bibinfo {author}
  {\bibfnamefont {T.}~\bibnamefont {Larsen}}, \bibinfo {author} {\bibfnamefont
  {L.~G.}\ \bibnamefont {Villanueva}},\ and\ \bibinfo {author} {\bibfnamefont
  {S.~D.}\ \bibnamefont {Huber}},\ }\bibfield  {title} {\bibinfo {title}
  {Observation of a phononic quadrupole topological insulator},\ }\href
  {https://doi.org/10.1038/nature25156} {\bibfield  {journal} {\bibinfo
  {journal} {Nature}\ }\textbf {\bibinfo {volume} {555}},\ \bibinfo {pages}
  {342} (\bibinfo {year} {2018})}\BibitemShut {NoStop}%
\bibitem [{\citenamefont {Qi}\ \emph {et~al.}(2020)\citenamefont {Qi},
  \citenamefont {Qiu}, \citenamefont {Xiao}, \citenamefont {He}, \citenamefont
  {Ke},\ and\ \citenamefont {Liu}}]{Acoustic2020}%
  \BibitemOpen
  \bibfield  {author} {\bibinfo {author} {\bibfnamefont {Y.}~\bibnamefont
  {Qi}}, \bibinfo {author} {\bibfnamefont {C.}~\bibnamefont {Qiu}}, \bibinfo
  {author} {\bibfnamefont {M.}~\bibnamefont {Xiao}}, \bibinfo {author}
  {\bibfnamefont {H.}~\bibnamefont {He}}, \bibinfo {author} {\bibfnamefont
  {M.}~\bibnamefont {Ke}},\ and\ \bibinfo {author} {\bibfnamefont
  {Z.}~\bibnamefont {Liu}},\ }\bibfield  {title} {\bibinfo {title} {Acoustic
  realization of quadrupole topological insulators},\ }\href
  {https://doi.org/10.1103/PhysRevLett.124.206601} {\bibfield  {journal}
  {\bibinfo  {journal} {Phys. Rev. Lett.}\ }\textbf {\bibinfo {volume} {124}},\
  \bibinfo {pages} {206601} (\bibinfo {year} {2020})}\BibitemShut {NoStop}%
\bibitem [{\citenamefont {Peterson}\ \emph {et~al.}(2018)\citenamefont
  {Peterson}, \citenamefont {Benalcazar}, \citenamefont {Hughes},\ and\
  \citenamefont {Bahl}}]{microwave_circuit2018}%
  \BibitemOpen
  \bibfield  {author} {\bibinfo {author} {\bibfnamefont {C.~W.}\ \bibnamefont
  {Peterson}}, \bibinfo {author} {\bibfnamefont {W.~A.}\ \bibnamefont
  {Benalcazar}}, \bibinfo {author} {\bibfnamefont {T.~L.}\ \bibnamefont
  {Hughes}},\ and\ \bibinfo {author} {\bibfnamefont {G.}~\bibnamefont {Bahl}},\
  }\bibfield  {title} {\bibinfo {title} {A quantized microwave quadrupole
  insulator with topologically protected corner states},\ }\href
  {https://doi.org/10.1038/nature25777} {\bibfield  {journal} {\bibinfo
  {journal} {Nature}\ }\textbf {\bibinfo {volume} {555}},\ \bibinfo {pages}
  {346} (\bibinfo {year} {2018})}\BibitemShut {NoStop}%
\bibitem [{\citenamefont {Imhof}\ \emph {et~al.}(2018)\citenamefont {Imhof},
  \citenamefont {Berger}, \citenamefont {Bayer}, \citenamefont {Brehm},
  \citenamefont {Molenkamp}, \citenamefont {Kiessling}, \citenamefont
  {Schindler}, \citenamefont {Lee}, \citenamefont {Greiter}, \citenamefont
  {Neupert},\ and\ \citenamefont {Thomale}}]{circuit2018}%
  \BibitemOpen
  \bibfield  {author} {\bibinfo {author} {\bibfnamefont {S.}~\bibnamefont
  {Imhof}}, \bibinfo {author} {\bibfnamefont {C.}~\bibnamefont {Berger}},
  \bibinfo {author} {\bibfnamefont {F.}~\bibnamefont {Bayer}}, \bibinfo
  {author} {\bibfnamefont {J.}~\bibnamefont {Brehm}}, \bibinfo {author}
  {\bibfnamefont {L.~W.}\ \bibnamefont {Molenkamp}}, \bibinfo {author}
  {\bibfnamefont {T.}~\bibnamefont {Kiessling}}, \bibinfo {author}
  {\bibfnamefont {F.}~\bibnamefont {Schindler}}, \bibinfo {author}
  {\bibfnamefont {C.~H.}\ \bibnamefont {Lee}}, \bibinfo {author} {\bibfnamefont
  {M.}~\bibnamefont {Greiter}}, \bibinfo {author} {\bibfnamefont
  {T.}~\bibnamefont {Neupert}},\ and\ \bibinfo {author} {\bibfnamefont
  {R.}~\bibnamefont {Thomale}},\ }\bibfield  {title} {\bibinfo {title}
  {Topolectrical-circuit realization of topological corner modes},\ }\href
  {https://doi.org/10.1038/s41567-018-0246-1} {\bibfield  {journal} {\bibinfo
  {journal} {Nature Physics}\ }\textbf {\bibinfo {volume} {14}},\ \bibinfo
  {pages} {925} (\bibinfo {year} {2018})}\BibitemShut {NoStop}%
\bibitem [{\citenamefont {Serra-Garcia}\ \emph {et~al.}(2019)\citenamefont
  {Serra-Garcia}, \citenamefont {S\"usstrunk},\ and\ \citenamefont
  {Huber}}]{circuit2019}%
  \BibitemOpen
  \bibfield  {author} {\bibinfo {author} {\bibfnamefont {M.}~\bibnamefont
  {Serra-Garcia}}, \bibinfo {author} {\bibfnamefont {R.}~\bibnamefont
  {S\"usstrunk}},\ and\ \bibinfo {author} {\bibfnamefont {S.~D.}\ \bibnamefont
  {Huber}},\ }\bibfield  {title} {\bibinfo {title} {Observation of quadrupole
  transitions and edge mode topology in an lc circuit network},\ }\href
  {https://doi.org/10.1103/PhysRevB.99.020304} {\bibfield  {journal} {\bibinfo
  {journal} {Phys. Rev. B}\ }\textbf {\bibinfo {volume} {99}},\ \bibinfo
  {pages} {020304} (\bibinfo {year} {2019})}\BibitemShut {NoStop}%
\bibitem [{\citenamefont {Bouhon}\ \emph {et~al.}(2019)\citenamefont {Bouhon},
  \citenamefont {Black-Schaffer},\ and\ \citenamefont {Slager}}]{Wilson2019}%
  \BibitemOpen
  \bibfield  {author} {\bibinfo {author} {\bibfnamefont {A.}~\bibnamefont
  {Bouhon}}, \bibinfo {author} {\bibfnamefont {A.~M.}\ \bibnamefont
  {Black-Schaffer}},\ and\ \bibinfo {author} {\bibfnamefont {R.-J.}\
  \bibnamefont {Slager}},\ }\bibfield  {title} {\bibinfo {title} {Wilson loop
  approach to fragile topology of split elementary band representations and
  topological crystalline insulators with time-reversal symmetry},\ }\href
  {https://doi.org/10.1103/PhysRevB.100.195135} {\bibfield  {journal} {\bibinfo
   {journal} {Phys. Rev. B}\ }\textbf {\bibinfo {volume} {100}},\ \bibinfo
  {pages} {195135} (\bibinfo {year} {2019})}\BibitemShut {NoStop}%
\bibitem [{\citenamefont {Li}\ \emph {et~al.}(2020)\citenamefont {Li},
  \citenamefont {Fu}, \citenamefont {Hu}, \citenamefont {Li},\ and\
  \citenamefont {Shen}}]{2020DisEQI}%
  \BibitemOpen
  \bibfield  {author} {\bibinfo {author} {\bibfnamefont {C.-A.}\ \bibnamefont
  {Li}}, \bibinfo {author} {\bibfnamefont {B.}~\bibnamefont {Fu}}, \bibinfo
  {author} {\bibfnamefont {Z.-A.}\ \bibnamefont {Hu}}, \bibinfo {author}
  {\bibfnamefont {J.}~\bibnamefont {Li}},\ and\ \bibinfo {author}
  {\bibfnamefont {S.-Q.}\ \bibnamefont {Shen}},\ }\bibfield  {title} {\bibinfo
  {title} {Topological phase transitions in disordered electric quadrupole
  insulators},\ }\href {https://doi.org/10.1103/PhysRevLett.125.166801}
  {\bibfield  {journal} {\bibinfo  {journal} {Phys. Rev. Lett.}\ }\textbf
  {\bibinfo {volume} {125}},\ \bibinfo {pages} {166801} (\bibinfo {year}
  {2020})}\BibitemShut {NoStop}%
\bibitem [{\citenamefont {Ren}\ \emph {et~al.}(2021)\citenamefont {Ren},
  \citenamefont {Souza},\ and\ \citenamefont
  {Vanderbilt}}]{Quadrupole_Wannier2021}%
  \BibitemOpen
  \bibfield  {author} {\bibinfo {author} {\bibfnamefont {S.}~\bibnamefont
  {Ren}}, \bibinfo {author} {\bibfnamefont {I.}~\bibnamefont {Souza}},\ and\
  \bibinfo {author} {\bibfnamefont {D.}~\bibnamefont {Vanderbilt}},\ }\bibfield
   {title} {\bibinfo {title} {Quadrupole moments, edge polarizations, and
  corner charges in the wannier representation},\ }\href
  {https://doi.org/10.1103/PhysRevB.103.035147} {\bibfield  {journal} {\bibinfo
   {journal} {Phys. Rev. B}\ }\textbf {\bibinfo {volume} {103}},\ \bibinfo
  {pages} {035147} (\bibinfo {year} {2021})}\BibitemShut {NoStop}%
\bibitem [{\citenamefont {Marzari}\ and\ \citenamefont
  {Vanderbilt}(1997)}]{localizedWannier1997}%
  \BibitemOpen
  \bibfield  {author} {\bibinfo {author} {\bibfnamefont {N.}~\bibnamefont
  {Marzari}}\ and\ \bibinfo {author} {\bibfnamefont {D.}~\bibnamefont
  {Vanderbilt}},\ }\bibfield  {title} {\bibinfo {title} {Maximally localized
  generalized wannier functions for composite energy bands},\ }\href
  {https://doi.org/10.1103/PhysRevB.56.12847} {\bibfield  {journal} {\bibinfo
  {journal} {Phys. Rev. B}\ }\textbf {\bibinfo {volume} {56}},\ \bibinfo
  {pages} {12847} (\bibinfo {year} {1997})}\BibitemShut {NoStop}%
\bibitem [{\citenamefont {Resta}(2011)}]{Resta2011}%
  \BibitemOpen
  \bibfield  {author} {\bibinfo {author} {\bibfnamefont {R.}~\bibnamefont
  {Resta}},\ }\bibfield  {title} {\bibinfo {title} {The insulating state of
  matter: a geometrical theory},\ }\href
  {https://doi.org/10.1140/epjb/e2010-10874-4} {\bibfield  {journal} {\bibinfo
  {journal} {The European Physical Journal B}\ }\textbf {\bibinfo {volume}
  {79}},\ \bibinfo {pages} {121} (\bibinfo {year} {2011})}\BibitemShut
  {NoStop}%
\bibitem [{\citenamefont {Marzari}\ \emph {et~al.}(2012)\citenamefont
  {Marzari}, \citenamefont {Mostofi}, \citenamefont {Yates}, \citenamefont
  {Souza},\ and\ \citenamefont {Vanderbilt}}]{RMPVanderbilt2012}%
  \BibitemOpen
  \bibfield  {author} {\bibinfo {author} {\bibfnamefont {N.}~\bibnamefont
  {Marzari}}, \bibinfo {author} {\bibfnamefont {A.~A.}\ \bibnamefont
  {Mostofi}}, \bibinfo {author} {\bibfnamefont {J.~R.}\ \bibnamefont {Yates}},
  \bibinfo {author} {\bibfnamefont {I.}~\bibnamefont {Souza}},\ and\ \bibinfo
  {author} {\bibfnamefont {D.}~\bibnamefont {Vanderbilt}},\ }\bibfield  {title}
  {\bibinfo {title} {Maximally localized wannier functions: Theory and
  applications},\ }\href {https://doi.org/10.1103/RevModPhys.84.1419}
  {\bibfield  {journal} {\bibinfo  {journal} {Rev. Mod. Phys.}\ }\textbf
  {\bibinfo {volume} {84}},\ \bibinfo {pages} {1419} (\bibinfo {year}
  {2012})}\BibitemShut {NoStop}%
\bibitem [{\citenamefont {Daido}\ \emph {et~al.}(2020)\citenamefont {Daido},
  \citenamefont {Shitade},\ and\ \citenamefont {Yanase}}]{EQM2020}%
  \BibitemOpen
  \bibfield  {author} {\bibinfo {author} {\bibfnamefont {A.}~\bibnamefont
  {Daido}}, \bibinfo {author} {\bibfnamefont {A.}~\bibnamefont {Shitade}},\
  and\ \bibinfo {author} {\bibfnamefont {Y.}~\bibnamefont {Yanase}},\
  }\bibfield  {title} {\bibinfo {title} {Thermodynamic approach to electric
  quadrupole moments},\ }\href {https://doi.org/10.1103/PhysRevB.102.235149}
  {\bibfield  {journal} {\bibinfo  {journal} {Phys. Rev. B}\ }\textbf {\bibinfo
  {volume} {102}},\ \bibinfo {pages} {235149} (\bibinfo {year}
  {2020})}\BibitemShut {NoStop}%
\bibitem [{\citenamefont {Kitamura}\ \emph {et~al.}(2021)\citenamefont
  {Kitamura}, \citenamefont {Ishizuka}, \citenamefont {Daido},\ and\
  \citenamefont {Yanase}}]{EQM2021}%
  \BibitemOpen
  \bibfield  {author} {\bibinfo {author} {\bibfnamefont {T.}~\bibnamefont
  {Kitamura}}, \bibinfo {author} {\bibfnamefont {J.}~\bibnamefont {Ishizuka}},
  \bibinfo {author} {\bibfnamefont {A.}~\bibnamefont {Daido}},\ and\ \bibinfo
  {author} {\bibfnamefont {Y.}~\bibnamefont {Yanase}},\ }\bibfield  {title}
  {\bibinfo {title} {Thermodynamic electric quadrupole moments of nematic
  phases from first-principles calculations},\ }\href
  {https://doi.org/10.1103/PhysRevB.103.245114} {\bibfield  {journal} {\bibinfo
   {journal} {Phys. Rev. B}\ }\textbf {\bibinfo {volume} {103}},\ \bibinfo
  {pages} {245114} (\bibinfo {year} {2021})}\BibitemShut {NoStop}%
\bibitem [{\citenamefont {Ono}\ \emph {et~al.}(2019)\citenamefont {Ono},
  \citenamefont {Trifunovic},\ and\ \citenamefont {Watanabe}}]{Watanabe2019}%
  \BibitemOpen
  \bibfield  {author} {\bibinfo {author} {\bibfnamefont {S.}~\bibnamefont
  {Ono}}, \bibinfo {author} {\bibfnamefont {L.}~\bibnamefont {Trifunovic}},\
  and\ \bibinfo {author} {\bibfnamefont {H.}~\bibnamefont {Watanabe}},\
  }\bibfield  {title} {\bibinfo {title} {Difficulties in operator-based
  formulation of the bulk quadrupole moment},\ }\href
  {https://doi.org/10.1103/PhysRevB.100.245133} {\bibfield  {journal} {\bibinfo
   {journal} {Phys. Rev. B}\ }\textbf {\bibinfo {volume} {100}},\ \bibinfo
  {pages} {245133} (\bibinfo {year} {2019})}\BibitemShut {NoStop}%
\bibitem [{\citenamefont {Watanabe}\ and\ \citenamefont
  {Ono}(2020)}]{Watanabe2020}%
  \BibitemOpen
  \bibfield  {author} {\bibinfo {author} {\bibfnamefont {H.}~\bibnamefont
  {Watanabe}}\ and\ \bibinfo {author} {\bibfnamefont {S.}~\bibnamefont {Ono}},\
  }\bibfield  {title} {\bibinfo {title} {Corner charge and bulk multipole
  moment in periodic systems},\ }\href
  {https://doi.org/10.1103/PhysRevB.102.165120} {\bibfield  {journal} {\bibinfo
   {journal} {Phys. Rev. B}\ }\textbf {\bibinfo {volume} {102}},\ \bibinfo
  {pages} {165120} (\bibinfo {year} {2020})}\BibitemShut {NoStop}%
\bibitem [{\citenamefont {Okugawa}\ \emph {et~al.}(2019)\citenamefont
  {Okugawa}, \citenamefont {Hayashi},\ and\ \citenamefont
  {Nakanishi}}]{SecondChiral2019}%
  \BibitemOpen
  \bibfield  {author} {\bibinfo {author} {\bibfnamefont {R.}~\bibnamefont
  {Okugawa}}, \bibinfo {author} {\bibfnamefont {S.}~\bibnamefont {Hayashi}},\
  and\ \bibinfo {author} {\bibfnamefont {T.}~\bibnamefont {Nakanishi}},\
  }\bibfield  {title} {\bibinfo {title} {Second-order topological phases
  protected by chiral symmetry},\ }\href
  {https://doi.org/10.1103/PhysRevB.100.235302} {\bibfield  {journal} {\bibinfo
   {journal} {Phys. Rev. B}\ }\textbf {\bibinfo {volume} {100}},\ \bibinfo
  {pages} {235302} (\bibinfo {year} {2019})}\BibitemShut {NoStop}%
\bibitem [{\citenamefont {Benalcazar}\ and\ \citenamefont
  {Cerjan}(2022)}]{benalcazar2022chiral}%
  \BibitemOpen
  \bibfield  {author} {\bibinfo {author} {\bibfnamefont {W.~A.}\ \bibnamefont
  {Benalcazar}}\ and\ \bibinfo {author} {\bibfnamefont {A.}~\bibnamefont
  {Cerjan}},\ }\bibfield  {title} {\bibinfo {title} {Chiral-symmetric
  higher-order topological phases of matter},\ }\href
  {https://doi.org/10.1103/PhysRevLett.128.127601} {\bibfield  {journal}
  {\bibinfo  {journal} {Phys. Rev. Lett.}\ }\textbf {\bibinfo {volume} {128}},\
  \bibinfo {pages} {127601} (\bibinfo {year} {2022})}\BibitemShut {NoStop}%
\bibitem [{SM()}]{SM}%
  \BibitemOpen
  \href@noop {} {\bibinfo {title} {See supplemental material at
  http://link.aps.org/supplemental/ 10.1103/physrevb.107.075413 for details of
  the calculations.}}\BibitemShut {Stop}%
\bibitem [{\citenamefont {Mondragon-Shem}\ \emph {et~al.}(2014)\citenamefont
  {Mondragon-Shem}, \citenamefont {Hughes}, \citenamefont {Song},\ and\
  \citenamefont {Prodan}}]{A3class2014}%
  \BibitemOpen
  \bibfield  {author} {\bibinfo {author} {\bibfnamefont {I.}~\bibnamefont
  {Mondragon-Shem}}, \bibinfo {author} {\bibfnamefont {T.~L.}\ \bibnamefont
  {Hughes}}, \bibinfo {author} {\bibfnamefont {J.}~\bibnamefont {Song}},\ and\
  \bibinfo {author} {\bibfnamefont {E.}~\bibnamefont {Prodan}},\ }\bibfield
  {title} {\bibinfo {title} {Topological criticality in the chiral-symmetric
  aiii class at strong disorder},\ }\href
  {https://doi.org/10.1103/PhysRevLett.113.046802} {\bibfield  {journal}
  {\bibinfo  {journal} {Phys. Rev. Lett.}\ }\textbf {\bibinfo {volume} {113}},\
  \bibinfo {pages} {046802} (\bibinfo {year} {2014})}\BibitemShut {NoStop}%
\bibitem [{\citenamefont {Asbóth}\ \emph {et~al.}(2016)\citenamefont
  {Asbóth}, \citenamefont {Oroszlány},\ and\ \citenamefont
  {Pályi}}]{ShortTopological}%
  \BibitemOpen
  \bibfield  {author} {\bibinfo {author} {\bibfnamefont {J.~K.}\ \bibnamefont
  {Asbóth}}, \bibinfo {author} {\bibfnamefont {L.}~\bibnamefont
  {Oroszlány}},\ and\ \bibinfo {author} {\bibfnamefont {A.}~\bibnamefont
  {Pályi}},\ }\href@noop {} {\emph {\bibinfo {title} {A Short Course on
  Topological Insulators}}},\ Lecture Notes in Physics\ (\bibinfo  {publisher}
  {Springer, Cham},\ \bibinfo {address} {Switzerland},\ \bibinfo {year}
  {2016})\BibitemShut {NoStop}%
\bibitem [{\citenamefont {Resta}(1994)}]{RMP_Resta1994}%
  \BibitemOpen
  \bibfield  {author} {\bibinfo {author} {\bibfnamefont {R.}~\bibnamefont
  {Resta}},\ }\bibfield  {title} {\bibinfo {title} {Macroscopic polarization in
  crystalline dielectrics: the geometric phase approach},\ }\href
  {https://doi.org/10.1103/RevModPhys.66.899} {\bibfield  {journal} {\bibinfo
  {journal} {Rev. Mod. Phys.}\ }\textbf {\bibinfo {volume} {66}},\ \bibinfo
  {pages} {899} (\bibinfo {year} {1994})}\BibitemShut {NoStop}%
\bibitem [{\citenamefont {Liu}\ and\ \citenamefont
  {Wakabayashi}(2017)}]{2DSSH2017}%
  \BibitemOpen
  \bibfield  {author} {\bibinfo {author} {\bibfnamefont {F.}~\bibnamefont
  {Liu}}\ and\ \bibinfo {author} {\bibfnamefont {K.}~\bibnamefont
  {Wakabayashi}},\ }\bibfield  {title} {\bibinfo {title} {Novel topological
  phase with a zero berry curvature},\ }\href
  {https://doi.org/10.1103/PhysRevLett.118.076803} {\bibfield  {journal}
  {\bibinfo  {journal} {Phys. Rev. Lett.}\ }\textbf {\bibinfo {volume} {118}},\
  \bibinfo {pages} {076803} (\bibinfo {year} {2017})}\BibitemShut {NoStop}%
\bibitem [{\citenamefont {Queiroz}\ \emph {et~al.}(2019)\citenamefont
  {Queiroz}, \citenamefont {Fulga}, \citenamefont {Avraham}, \citenamefont
  {Beidenkopf},\ and\ \citenamefont {Cano}}]{Partial2019}%
  \BibitemOpen
  \bibfield  {author} {\bibinfo {author} {\bibfnamefont {R.}~\bibnamefont
  {Queiroz}}, \bibinfo {author} {\bibfnamefont {I.~C.}\ \bibnamefont {Fulga}},
  \bibinfo {author} {\bibfnamefont {N.}~\bibnamefont {Avraham}}, \bibinfo
  {author} {\bibfnamefont {H.}~\bibnamefont {Beidenkopf}},\ and\ \bibinfo
  {author} {\bibfnamefont {J.}~\bibnamefont {Cano}},\ }\bibfield  {title}
  {\bibinfo {title} {Partial lattice defects in higher-order topological
  insulators},\ }\href {https://doi.org/10.1103/PhysRevLett.123.266802}
  {\bibfield  {journal} {\bibinfo  {journal} {Phys. Rev. Lett.}\ }\textbf
  {\bibinfo {volume} {123}},\ \bibinfo {pages} {266802} (\bibinfo {year}
  {2019})}\BibitemShut {NoStop}%
\bibitem [{\citenamefont {Li}\ and\ \citenamefont
  {Miroshnichenko}(2019)}]{ExtendedSSH}%
  \BibitemOpen
  \bibfield  {author} {\bibinfo {author} {\bibfnamefont {C.}~\bibnamefont
  {Li}}\ and\ \bibinfo {author} {\bibfnamefont {A.~E.}\ \bibnamefont
  {Miroshnichenko}},\ }\bibfield  {title} {\bibinfo {title} {Extended ssh
  model: Non-local couplings and non-monotonous edge states},\ }\href
  {https://doi.org/10.3390/physics1010002} {\bibfield  {journal} {\bibinfo
  {journal} {Physics}\ }\textbf {\bibinfo {volume} {1}},\ \bibinfo {pages} {2}
  (\bibinfo {year} {2019})}\BibitemShut {NoStop}%
\bibitem [{\citenamefont {Hayashi}(2019)}]{Shin2019}%
  \BibitemOpen
  \bibfield  {author} {\bibinfo {author} {\bibfnamefont {S.}~\bibnamefont
  {Hayashi}},\ }\bibfield  {title} {\bibinfo {title} {Toeplitz operators on
  concave corners and topologically protected corner states},\ }\href
  {https://doi.org/10.1007/s11005-019-01184-w} {\bibfield  {journal} {\bibinfo
  {journal} {Letters in Mathematical Physics}\ }\textbf {\bibinfo {volume}
  {109}},\ \bibinfo {pages} {2223} (\bibinfo {year} {2019})}\BibitemShut
  {NoStop}%
\bibitem [{\citenamefont {Hayashi}(2021)}]{Shin2021}%
  \BibitemOpen
  \bibfield  {author} {\bibinfo {author} {\bibfnamefont {S.}~\bibnamefont
  {Hayashi}},\ }\bibfield  {title} {\bibinfo {title} {Classification of
  topological invariants related to corner states},\ }\href
  {https://doi.org/10.1007/s11005-021-01460-8} {\bibfield  {journal} {\bibinfo
  {journal} {Letters in Mathematical Physics}\ }\textbf {\bibinfo {volume}
  {111}},\ \bibinfo {pages} {118} (\bibinfo {year} {2021})}\BibitemShut
  {NoStop}%
\bibitem [{\citenamefont {Franca}\ \emph {et~al.}(2018)\citenamefont {Franca},
  \citenamefont {van~den Brink},\ and\ \citenamefont
  {Fulga}}]{anomalousHOTs2018}%
  \BibitemOpen
  \bibfield  {author} {\bibinfo {author} {\bibfnamefont {S.}~\bibnamefont
  {Franca}}, \bibinfo {author} {\bibfnamefont {J.}~\bibnamefont {van~den
  Brink}},\ and\ \bibinfo {author} {\bibfnamefont {I.~C.}\ \bibnamefont
  {Fulga}},\ }\bibfield  {title} {\bibinfo {title} {An anomalous higher-order
  topological insulator},\ }\href {https://doi.org/10.1103/PhysRevB.98.201114}
  {\bibfield  {journal} {\bibinfo  {journal} {Phys. Rev. B}\ }\textbf {\bibinfo
  {volume} {98}},\ \bibinfo {pages} {201114} (\bibinfo {year}
  {2018})}\BibitemShut {NoStop}%
\bibitem [{\citenamefont {Souza}\ \emph {et~al.}(2000)\citenamefont {Souza},
  \citenamefont {Wilkens},\ and\ \citenamefont {Martin}}]{geometry_metric2000}%
  \BibitemOpen
  \bibfield  {author} {\bibinfo {author} {\bibfnamefont {I.}~\bibnamefont
  {Souza}}, \bibinfo {author} {\bibfnamefont {T.}~\bibnamefont {Wilkens}},\
  and\ \bibinfo {author} {\bibfnamefont {R.~M.}\ \bibnamefont {Martin}},\
  }\bibfield  {title} {\bibinfo {title} {Polarization and localization in
  insulators: Generating function approach},\ }\href
  {https://doi.org/10.1103/PhysRevB.62.1666} {\bibfield  {journal} {\bibinfo
  {journal} {Phys. Rev. B}\ }\textbf {\bibinfo {volume} {62}},\ \bibinfo
  {pages} {1666} (\bibinfo {year} {2000})}\BibitemShut {NoStop}%
\bibitem [{\citenamefont {Bleu}\ \emph {et~al.}(2018)\citenamefont {Bleu},
  \citenamefont {Malpuech}, \citenamefont {Gao},\ and\ \citenamefont
  {Solnyshkov}}]{GMT2018}%
  \BibitemOpen
  \bibfield  {author} {\bibinfo {author} {\bibfnamefont {O.}~\bibnamefont
  {Bleu}}, \bibinfo {author} {\bibfnamefont {G.}~\bibnamefont {Malpuech}},
  \bibinfo {author} {\bibfnamefont {Y.}~\bibnamefont {Gao}},\ and\ \bibinfo
  {author} {\bibfnamefont {D.~D.}\ \bibnamefont {Solnyshkov}},\ }\bibfield
  {title} {\bibinfo {title} {Effective theory of nonadiabatic quantum evolution
  based on the quantum geometric tensor},\ }\href
  {https://doi.org/10.1103/PhysRevLett.121.020401} {\bibfield  {journal}
  {\bibinfo  {journal} {Phys. Rev. Lett.}\ }\textbf {\bibinfo {volume} {121}},\
  \bibinfo {pages} {020401} (\bibinfo {year} {2018})}\BibitemShut {NoStop}%
\bibitem [{\citenamefont {Tan}\ \emph {et~al.}(2019)\citenamefont {Tan},
  \citenamefont {Zhang}, \citenamefont {Yang}, \citenamefont {Chu},
  \citenamefont {Zhu}, \citenamefont {Li}, \citenamefont {Yang}, \citenamefont
  {Song}, \citenamefont {Han}, \citenamefont {Li}, \citenamefont {Dong},
  \citenamefont {Yu}, \citenamefont {Yan}, \citenamefont {Zhu},\ and\
  \citenamefont {Yu}}]{QuantumMetric2019}%
  \BibitemOpen
  \bibfield  {author} {\bibinfo {author} {\bibfnamefont {X.}~\bibnamefont
  {Tan}}, \bibinfo {author} {\bibfnamefont {D.-W.}\ \bibnamefont {Zhang}},
  \bibinfo {author} {\bibfnamefont {Z.}~\bibnamefont {Yang}}, \bibinfo {author}
  {\bibfnamefont {J.}~\bibnamefont {Chu}}, \bibinfo {author} {\bibfnamefont
  {Y.-Q.}\ \bibnamefont {Zhu}}, \bibinfo {author} {\bibfnamefont
  {D.}~\bibnamefont {Li}}, \bibinfo {author} {\bibfnamefont {X.}~\bibnamefont
  {Yang}}, \bibinfo {author} {\bibfnamefont {S.}~\bibnamefont {Song}}, \bibinfo
  {author} {\bibfnamefont {Z.}~\bibnamefont {Han}}, \bibinfo {author}
  {\bibfnamefont {Z.}~\bibnamefont {Li}}, \bibinfo {author} {\bibfnamefont
  {Y.}~\bibnamefont {Dong}}, \bibinfo {author} {\bibfnamefont {H.-F.}\
  \bibnamefont {Yu}}, \bibinfo {author} {\bibfnamefont {H.}~\bibnamefont
  {Yan}}, \bibinfo {author} {\bibfnamefont {S.-L.}\ \bibnamefont {Zhu}},\ and\
  \bibinfo {author} {\bibfnamefont {Y.}~\bibnamefont {Yu}},\ }\bibfield
  {title} {\bibinfo {title} {Experimental measurement of the quantum metric
  tensor and related topological phase transition with a superconducting
  qubit},\ }\href {https://doi.org/10.1103/PhysRevLett.122.210401} {\bibfield
  {journal} {\bibinfo  {journal} {Phys. Rev. Lett.}\ }\textbf {\bibinfo
  {volume} {122}},\ \bibinfo {pages} {210401} (\bibinfo {year}
  {2019})}\BibitemShut {NoStop}%
\end{thebibliography}%


\begin{thebibliography}{9}%
\makeatletter
\providecommand \@ifxundefined [1]{%
 \@ifx{#1\undefined}
}%
\providecommand \@ifnum [1]{%
 \ifnum #1\expandafter \@firstoftwo
 \else \expandafter \@secondoftwo
 \fi
}%
\providecommand \@ifx [1]{%
 \ifx #1\expandafter \@firstoftwo
 \else \expandafter \@secondoftwo
 \fi
}%
\providecommand \natexlab [1]{#1}%
\providecommand \enquote  [1]{``#1''}%
\providecommand \bibnamefont  [1]{#1}%
\providecommand \bibfnamefont [1]{#1}%
\providecommand \citenamefont [1]{#1}%
\providecommand \href@noop [0]{\@secondoftwo}%
\providecommand \href [0]{\begingroup \@sanitize@url \@href}%
\providecommand \@href[1]{\@@startlink{#1}\@@href}%
\providecommand \@@href[1]{\endgroup#1\@@endlink}%
\providecommand \@sanitize@url [0]{\catcode `\\12\catcode `\$12\catcode
  `\&12\catcode `\#12\catcode `\^12\catcode `\_12\catcode `\%12\relax}%
\providecommand \@@startlink[1]{}%
\providecommand \@@endlink[0]{}%
\providecommand \url  [0]{\begingroup\@sanitize@url \@url }%
\providecommand \@url [1]{\endgroup\@href {#1}{\urlprefix }}%
\providecommand \urlprefix  [0]{URL }%
\providecommand \Eprint [0]{\href }%
\providecommand \doibase [0]{https://doi.org/}%
\providecommand \selectlanguage [0]{\@gobble}%
\providecommand \bibinfo  [0]{\@secondoftwo}%
\providecommand \bibfield  [0]{\@secondoftwo}%
\providecommand \translation [1]{[#1]}%
\providecommand \BibitemOpen [0]{}%
\providecommand \bibitemStop [0]{}%
\providecommand \bibitemNoStop [0]{.\EOS\space}%
\providecommand \EOS [0]{\spacefactor3000\relax}%
\providecommand \BibitemShut  [1]{\csname bibitem#1\endcsname}%
\let\auto@bib@innerbib\@empty
\bibitem [{\citenamefont {Benalcazar}\ \emph
  {et~al.}(2017{\natexlab{a}})\citenamefont {Benalcazar}, \citenamefont
  {Bernevig},\ and\ \citenamefont {Hughes}}]{multipole_PRB_2017}%
  \BibitemOpen
  \bibfield  {author} {\bibinfo {author} {\bibfnamefont {W.~A.}\ \bibnamefont
  {Benalcazar}}, \bibinfo {author} {\bibfnamefont {B.~A.}\ \bibnamefont
  {Bernevig}},\ and\ \bibinfo {author} {\bibfnamefont {T.~L.}\ \bibnamefont
  {Hughes}},\ }\bibfield  {title} {\bibinfo {title} {Electric multipole
  moments, topological multipole moment pumping, and chiral hinge states in
  crystalline insulators},\ }\href {https://doi.org/10.1103/PhysRevB.96.245115}
  {\bibfield  {journal} {\bibinfo  {journal} {Phys. Rev. B}\ }\textbf {\bibinfo
  {volume} {96}},\ \bibinfo {pages} {245115} (\bibinfo {year}
  {2017}{\natexlab{a}})}\BibitemShut {NoStop}%
\bibitem [{\citenamefont {Benalcazar}\ \emph
  {et~al.}(2017{\natexlab{b}})\citenamefont {Benalcazar}, \citenamefont
  {Bernevig},\ and\ \citenamefont {Hughes}}]{multipole_science2017}%
  \BibitemOpen
  \bibfield  {author} {\bibinfo {author} {\bibfnamefont {W.~A.}\ \bibnamefont
  {Benalcazar}}, \bibinfo {author} {\bibfnamefont {B.~A.}\ \bibnamefont
  {Bernevig}},\ and\ \bibinfo {author} {\bibfnamefont {T.~L.}\ \bibnamefont
  {Hughes}},\ }\bibfield  {title} {\bibinfo {title} {Quantized electric
  multipole insulators},\ }\href {https://doi.org/10.1126/science.aah6442}
  {\bibfield  {journal} {\bibinfo  {journal} {Science}\ }\textbf {\bibinfo
  {volume} {357}},\ \bibinfo {pages} {61–66} (\bibinfo {year}
  {2017}{\natexlab{b}})}\BibitemShut {NoStop}%
\bibitem [{\citenamefont {Resta}(1994)}]{RMP_Resta1994}%
  \BibitemOpen
  \bibfield  {author} {\bibinfo {author} {\bibfnamefont {R.}~\bibnamefont
  {Resta}},\ }\bibfield  {title} {\bibinfo {title} {Macroscopic polarization in
  crystalline dielectrics: the geometric phase approach},\ }\href
  {https://doi.org/10.1103/RevModPhys.66.899} {\bibfield  {journal} {\bibinfo
  {journal} {Rev. Mod. Phys.}\ }\textbf {\bibinfo {volume} {66}},\ \bibinfo
  {pages} {899} (\bibinfo {year} {1994})}\BibitemShut {NoStop}%
\bibitem [{\citenamefont {Liu}\ and\ \citenamefont
  {Wakabayashi}(2017)}]{2DSSH2017}%
  \BibitemOpen
  \bibfield  {author} {\bibinfo {author} {\bibfnamefont {F.}~\bibnamefont
  {Liu}}\ and\ \bibinfo {author} {\bibfnamefont {K.}~\bibnamefont
  {Wakabayashi}},\ }\bibfield  {title} {\bibinfo {title} {Novel topological
  phase with a zero berry curvature},\ }\href
  {https://doi.org/10.1103/PhysRevLett.118.076803} {\bibfield  {journal}
  {\bibinfo  {journal} {Phys. Rev. Lett.}\ }\textbf {\bibinfo {volume} {118}},\
  \bibinfo {pages} {076803} (\bibinfo {year} {2017})}\BibitemShut {NoStop}%
\bibitem [{\citenamefont {Li}\ and\ \citenamefont
  {Miroshnichenko}(2019)}]{ExtendedSSH}%
  \BibitemOpen
  \bibfield  {author} {\bibinfo {author} {\bibfnamefont {C.}~\bibnamefont
  {Li}}\ and\ \bibinfo {author} {\bibfnamefont {A.~E.}\ \bibnamefont
  {Miroshnichenko}},\ }\bibfield  {title} {\bibinfo {title} {Extended ssh
  model: Non-local couplings and non-monotonous edge states},\ }\href
  {https://doi.org/10.3390/physics1010002} {\bibfield  {journal} {\bibinfo
  {journal} {Physics}\ }\textbf {\bibinfo {volume} {1}},\ \bibinfo {pages} {2}
  (\bibinfo {year} {2019})}\BibitemShut {NoStop}%
\bibitem [{\citenamefont {Li}\ and\ \citenamefont {Wu}(2020)}]{Changan2020}%
  \BibitemOpen
  \bibfield  {author} {\bibinfo {author} {\bibfnamefont {C.-A.}\ \bibnamefont
  {Li}}\ and\ \bibinfo {author} {\bibfnamefont {S.-S.}\ \bibnamefont {Wu}},\
  }\bibfield  {title} {\bibinfo {title} {Topological states in generalized
  electric quadrupole insulators},\ }\href
  {https://doi.org/10.1103/PhysRevB.101.195309} {\bibfield  {journal} {\bibinfo
   {journal} {Phys. Rev. B}\ }\textbf {\bibinfo {volume} {101}},\ \bibinfo
  {pages} {195309} (\bibinfo {year} {2020})}\BibitemShut {NoStop}%
\bibitem [{\citenamefont {Resta}(2011)}]{Resta2011}%
  \BibitemOpen
  \bibfield  {author} {\bibinfo {author} {\bibfnamefont {R.}~\bibnamefont
  {Resta}},\ }\bibfield  {title} {\bibinfo {title} {The insulating state of
  matter: a geometrical theory},\ }\href
  {https://doi.org/10.1140/epjb/e2010-10874-4} {\bibfield  {journal} {\bibinfo
  {journal} {The European Physical Journal B}\ }\textbf {\bibinfo {volume}
  {79}},\ \bibinfo {pages} {121} (\bibinfo {year} {2011})}\BibitemShut
  {NoStop}%
\bibitem [{\citenamefont {Souza}\ \emph {et~al.}(2000)\citenamefont {Souza},
  \citenamefont {Wilkens},\ and\ \citenamefont {Martin}}]{geometry_metric2000}%
  \BibitemOpen
  \bibfield  {author} {\bibinfo {author} {\bibfnamefont {I.}~\bibnamefont
  {Souza}}, \bibinfo {author} {\bibfnamefont {T.}~\bibnamefont {Wilkens}},\
  and\ \bibinfo {author} {\bibfnamefont {R.~M.}\ \bibnamefont {Martin}},\
  }\bibfield  {title} {\bibinfo {title} {Polarization and localization in
  insulators: Generating function approach},\ }\href
  {https://doi.org/10.1103/PhysRevB.62.1666} {\bibfield  {journal} {\bibinfo
  {journal} {Phys. Rev. B}\ }\textbf {\bibinfo {volume} {62}},\ \bibinfo
  {pages} {1666} (\bibinfo {year} {2000})}\BibitemShut {NoStop}%
\bibitem [{\citenamefont {Ono}\ \emph {et~al.}(2019)\citenamefont {Ono},
  \citenamefont {Trifunovic},\ and\ \citenamefont {Watanabe}}]{Watanabe2019}%
  \BibitemOpen
  \bibfield  {author} {\bibinfo {author} {\bibfnamefont {S.}~\bibnamefont
  {Ono}}, \bibinfo {author} {\bibfnamefont {L.}~\bibnamefont {Trifunovic}},\
  and\ \bibinfo {author} {\bibfnamefont {H.}~\bibnamefont {Watanabe}},\
  }\bibfield  {title} {\bibinfo {title} {Difficulties in operator-based
  formulation of the bulk quadrupole moment},\ }\href
  {https://doi.org/10.1103/PhysRevB.100.245133} {\bibfield  {journal} {\bibinfo
   {journal} {Phys. Rev. B}\ }\textbf {\bibinfo {volume} {100}},\ \bibinfo
  {pages} {245133} (\bibinfo {year} {2019})}\BibitemShut {NoStop}%
\end{thebibliography}%

\end{document}


\title{Supplementary Material for ``Wannier Topology and Quadrupole Moments for a generalized Benalcazar-Bernevig-Hughes Model"}
\author{Liu Yang} 
\email{liu.yang-2@manchester.ac.uk}
\affiliation{Department of Physics and Astronomy, The University of Manchester, Manchester M13 9PL, United Kingdom }
\author{Alessandro Principi}
\affiliation{Department of Physics and Astronomy, The University of Manchester, Manchester M13 9PL, United Kingdom }
\author{Niels R. Walet}
\affiliation{Department of Physics and Astronomy, The University of Manchester, Manchester M13 9PL, United Kingdom } %

\maketitle
In this Supplementary Material, we review the previous theoretical framework as given in Refs~\cite{multipole_PRB_2017,multipole_science2017} and show in detail the derivation of the main quantities discussed in the main text.
\section{Bulk dipole momentum in 1D crystal}\label{sec.1Ddipole}
In the second quantization, a generic 1D Hamiltonian with periodic boundary conditions can be written as
\begin{align}
\hat{H}=\sum_{k} c_{k, \alpha}^{\dagger}\left[h_{k}\right]_{\alpha, \beta} c_{k, \beta},
\end{align}
where $c_{k,\alpha}^\dagger$ is the electron creation operator for a state with momentum $k$ and orbital label $\alpha$. The energy eigen equation can be written as
\begin{align}
h(k)u_n(k)=E_n(k)u_n(k).
\end{align}
The dipole moment of a 1D crystal can be calculated through the Wilson loop, which is constructed through the multiplication of the exponents of the non-Abelian Berry connection
\begin{align}
\mathcal{W}_{k}=\lim_{\Delta_k\to 0}F_{2\pi-\Delta_k}F_{2\pi-2\Delta_k}\ldots F_{k+\Delta_k}F_{k},\label{Wilson_loop}
\end{align}
where $[F_k]_{n_1n_2}=u^\dagger_{n_1}(k+\Delta_k)u_{n_2}(k)$. For small $\Delta k$, $F_k$ can be approximated as
\begin{align}
F_k&\approx e^{i \Delta_k\mathcal{A}(k)},
\end{align}
where the non-Abelian Berry connection is defined as
\begin{align}
[\mathcal{A}(k)]_{n_1n_2}&=i   u^\dagger_{n_1}(k)\partial_ku_{n_1}(k).
\end{align}
We then solve for the eigenstates of the Wilson loop, 
\begin{align}
\mathcal{W}_{k}v_j(k)=\phi^j v_j(k)\quad,\forall\,k.
\end{align}
Since the Wilson loop is unitary, its eigenvalues can be expressed as a phase
\begin{align}
\phi^j=e^{i 2\pi\nu^j}.
\end{align}
In fact, $\nu^j$ is the Wannier center relative to the middle of the unit cells. In general, we call the eigenstates $v_j$ with its Wannier center $\nu^j$ as the Wannier bands. The corresponding classical polarization is defined as 
\begin{align}
p&=\sum_j(\nu^j\bmod 1)\nonumber\\
&=\frac{1}{2\pi i }\ln\det[\mathcal{W}_{k+2\pi\gets k}]\bmod 1\nonumber\\
&=\frac{1}{2\pi }\int_k^{k+2\pi}\Tr[\mathcal{A}(k)]\,\dd k\bmod 1.
\end{align}
Here $\int_k^{k+2\pi}\Tr[\mathcal{A}_k]\dd k$ is the Zak-Berry phase of all occupied bands, and we set the electron charge $\text{e}=1$. Note that the Zak-Berry phase may not be trivial when the polarization is trivial. 

We will use the following relation as a gauge-fixing condition for the eigenfunctions of the Wilson loop
\begin{align}
v_j(k+\Delta_k)&=e^{-i \Delta_k\nu^j}F_kv_j(k).\label{wannier_gauge}
\end{align}
Note that the choice of this gauge here is only for the convenience of calculation. Since there are different choices of the sub-lattice degree of freedom in the occupied bands, we can mix the eigenstates in the occupied subspace as follow
\begin{align}
|u'_{m}(k)\rangle=\sum_{n=1}^{N_{\text{occ}}}|u_n(k)\rangle[U_k]^{nm}.
\end{align}
This transformation makes the connection change as
\begin{align}
\mathcal{A}_k^{'}=iU^\dag_k\partial_kU_k-U^\dag_k\mathcal{A}_kU_k.
\end{align}
This gives a new polarization as
\begin{align}
p^{'}&=p+\frac{i }{2\pi}\text{Tr[ln} U_k]\bigg|_{k}^{k+2\pi}\nonumber\\
&=p+\frac{i }{2\pi}\text{ln[det} U_k]\bigg|_{k}^{k+2\pi}\nonumber\\
&=(p+n)\,\text{mod}\,1\nonumber\\
&=p.\label{1D_p_gauge}
\end{align}
In the above derivation, we apply the unitary property det$[U_k]=e^{-i\phi(k)}$ and $\phi(k+2\pi)=\phi(k)+2\pi n$ derived from periodic condition of $U_k$. Therefore, we can see that $p$ is a gauge-invariant quantity.

\section{Bulk Dipole moments and Zak-Berry phases in 2D crystal}
\label{sec:2D_dipole}
\subsection{formalism}

In two dimensions, we can extend the Wilson loop $\mathcal{W}_{x,\vec{k}}$ that runs along $k_x$ axis as a loop at a fixed value of $k_y$ and from the base point $\vec{k}$ in the 2D FBZ. With this extension, we have 
\begin{align}
\mathcal{W}_{x,\vec{k}}v_j^x(\vec{k})=\phi^j_xv_j^x(\vec{k}),
\end{align}
where
\begin{align}
\phi^j_{x}(k_y)=e^{i 2\pi\nu_x^j(k_y)}.
\end{align}
Then, the 1D polarization at fixed $k_y$ can be written as 
\begin{align}
p_x(k_y)=\sum_{j=1}^{N_{\text{occ}}}\nu_x^j(k_y)\bmod 1=-\frac{i }{2\pi}\ln\,\det[\mathcal{W}_{x,\vec{k}}]\bmod 1.
\end{align}
Thus the total polarization along $x$ is 
\begin{align}
p_x&=\frac{\Delta_{k_y}}{2\pi}\sum_{k_y}p_x(k_y)=\frac{1}{2\pi}\Omega^{2D}_y \bmod 1,\\
\Omega^{2D}_x&=\frac{1}{2\pi}\int_{\text{BZ}}\,\Tr[\mathcal{A}_{x}(\vec{k})]\,\dd^2\vec{k}.
\end{align}
By the same method, the total polarization along $y$ is 
\begin{align}
p_y&=\frac{\Delta_{k_x}}{2\pi}\sum_{k_x}p_y(k_x)=\frac{1}{2\pi}\Omega^{2D}_y\bmod 1,\\
\Omega^{2D}_y&=\frac{1}{2\pi}\int_{\text{BZ}}\,\Tr[\mathcal{A}_{y}(\vec{k})]\,\dd^2\vec{k}.
\end{align}
The functions $\Omega_\mu^{2D}$ ($\mu=x,y$) are called the 2D Zak-Berry phases\cite{RMP_Resta1994,2DSSH2017}.

\subsection{Polarizations in the generalized BBH model}
We first study a two-by-two Bloch Hamiltonian of the form 
  \begin{align}
  h=\vec{b}\cdot\vec{\sigma},
  \end{align}
where $\vec{\sigma}$ is the vector of  Pauli matrices $\vec{\sigma}=(s_1,s_2,s_3)$. Using spherical coordinates, we express the characteristic vector $\vec b$ as $\vec{b}=|\vec{b}|(\sin{\theta}\cos{\phi},\sin{\theta}\sin{\phi},\cos{\theta})$. Then the two eigenstates with eigenenergies $\pm|\vec{b}|$ can be written as
  \begin{align}
  \psi_+=\left(\begin{array}{c}
{ e^{-i \phi}\cos{\frac{\theta}{2}}} \\
{ \sin{\frac{\theta}{2}}} 
\end{array}\right),\quad\psi_-=\left(\begin{array}{c}
{ e^{-i \phi}\sin{\frac{\theta}{2}}} \\
{ -\cos{\frac{\theta}{2}}} 
\end{array}\right).\label{TwoBand}
  \end{align}

  The Bloch Hamiltonian of the generalized BBH model, as considered in the main text, is 
\begin{align}
\mathcal{H}(k_x,k_y)=h_x(k_x)\otimes\tau_3 +\sigma_0\otimes h_y(k_y),\label{hBBH}
\end{align}
where $h_x(k_x)=\vec{b}_x(k_x)\cdot\vec{\sigma}$, $h_y(k_y)=\vec{b}_y(k_y)\cdot\vec{\tau}$. The four sub-lattice sites in one unit cell are denoted as $\ket{\alpha_x\,\beta_y}$  ($\alpha,\beta=A\text{ or }B$). The eigenenergies of $\mathcal{H}(\vec{k})$ are $E_\pm(\vec{k})=\pm\sqrt{|\vec{b}_x(k_x)|^2+|\vec{b}_y(k_x)|^2}$. For the two degenerate-occupied bands, the orthogonal eigenstates are
\begin{align}
\Psi_{n}(k_x,k_y)&=\psi^x_{\sigma(n)}(k_x)\otimes\varphi^{n}_-(k_x,k_y).\label{Psin}
\end{align}
Here $n=1,2$ while $\sigma(n)=+,-$, $\psi^x_{\sigma(n)}(k_x)$ are the eigenstates of $h_x(k_x)$ with eigen-energies $\sigma(n) |\vec{b}_x(\vec{k})|$ and $\varphi^{n}_-(\vec{k})$ are the eigenstates with negative energies of the two-by-two Hamiltonians $h_{xy}^n(\vec{k})=\vec{b}^n_{xy}(\vec{k})\cdot\vec{\tau}$ whose characteristic vectors are $\vec{b}^n_{xy}(\vec{k})=(b^1_{y}(k_y),b^2_{y}(k_y),(-1)^{n-1}|\vec{b}_{x}(k_x)|)$, respectively. We then express the eigenstates $\psi^x_\pm(k_x)$ of $h_x(k_x)$ and $\varphi^{1/2}_\pm(k_x)$ of $h^{1/2}_{xy}(\vec{k})$ in spherical coordinates as follows
  \begin{align}
\psi^x_\pm(k_x)&=\frac{1}{\sqrt{2}}\left(\begin{array}{c}
{ e^{-i \phi_x(k_x)}} \\
{ \pm 1} 
\end{array}\right),\label{gaugepsi}\\
\varphi^1_-(k_x,k_y)&=\left(\begin{array}{c}
{ e^{-i \phi_y(k_y)}\sin{\frac{\theta_{xy}(\vec{k})}{2}}} \\
{ - \cos{\frac{\theta_{xy}(\vec{k})}{2}}} 
\end{array}\right),\label{gaugephi1}\\
\varphi^2_-(k_x,k_y)&=\left(\begin{array}{c}
{ e^{-i \phi_y(k_y)}\cos{\frac{\theta_{xy}(\vec{k})}{2}}} \\
{ - \sin{\frac{\theta_{xy}(\vec{k})}{2}}} 
\end{array}\right).\label{gaugephi2}
  \end{align}
  The angles in the above expressions are defined through the polar decomposition of the characteristic vectors $\vec{b}_x$ and $\vec{b}^{1}_{xy}=\vec{b}_y+(0,0,|\vec{b}_x|)$:
  \begin{align}
  \vec{b}_x(k_x)&=|\vec{b}_x(k_x)|\left(\cos{\phi_x(k_x)},\sin{\phi_x(k_x)},0\right),\\
   \vec{b}_{y}(k_y)&=|\vec{b}_y(k_y)|\left(\cos{\phi_y(k_y)},\sin{\phi_y(k_y)},0\right),\\
      \vec{b}^{1}_{xy}(k_x,k_y)&=(|\vec{b}_y(k_y)|\cos{\phi_y(k_y)},|\vec{b}_y(k_y)|\sin{\phi_y(k_y)},|\vec{b}_x(k_x)|)\nonumber\\
      &=\sqrt{ \vec{b}_x^2+\vec{b}_y^2}\,\left(\sin{\theta_{xy}(\vec{k})}\cos{\phi_y(k_y)},\sin{\theta_{xy}(\vec{k})}\sin{\phi_y(k_y)},\cos{\theta_{xy}(\vec{k})}\right).
  \end{align}
  Note that $\phi_\mu$ only depends on $k_\mu$ while $\theta_{xy}(\vec{k})$ is determined by both $k_x$ and $k_y$. More precisely,  
  \begin{align}
  \theta_{xy}(k_x,k_y)=\arccos{(|\vec{b}_x(k_x)|/\sqrt{ \vec{b}_x^2(k_x)+\vec{b}_y^2(k_y)})}.
  \end{align}
Note that, the gauge choices for the two-component spinors $\psi^x_{\sigma(n)}(k_x)$ and $\varphi^{n}_-(k_x,k_y)$ in Eqs.~(\ref{gaugepsi})-(\ref{gaugephi2}) are consistent with Eq.~(\ref{TwoBand}). The two eigenstates $\Psi_n(\vec{k})$ can be transformed into each other by
\begin{align}
\Psi_2(k_x,k_y)&=-e^{-i\phi_y}\sigma_3\otimes\tau_1\mathcal{K}\Psi_1(k_x,k_y).
\end{align}
 
 To obtain the bulk dipole moments of the generalized BBH model, we first calculate the diagonal entries of the non-Abelian Berry connections as follows
\begin{align}
[\mathcal{A}_{x}(\vec{k})]_{nn}&=i \Psi^\dagger_{n}(\vec{k})\partial_{k_x}\Psi_{n}(\vec{k})=i \psi_-^{x\dagger}(k_x)\partial_{k_x}\psi^x_-(k_x)+i \varphi_-^{n\dagger}(\vec{k})\partial_{k_x}\varphi^{n}_-(\vec{k}),\\
[\mathcal{A}_{y}(\vec{k})]_{nn}&=i \Psi_{n}^\dagger(\vec{k})\partial_{k_y}\Psi_{n}(\vec{k})=i \varphi_-^{n\dagger}(\vec{k})\partial_{k_y}\varphi^{n}_-(\vec{k}).
\end{align}
Here we have applied the fact that the two terms $i \psi^{x\dagger}_\pm(k_x)\partial_{k_x}\psi^x_\pm(k_x)$ are actually the same as $\psi^x_+(k_x)=\sigma_3\psi^x_-(k_x)$, which originates from the chiral symmetry. Applying
\begin{align}
\varphi^{2}_-(\vec{k})=-e^{-i\phi_y}\tau_1\mathcal{K}\varphi^{1}_-(\vec{k}),
\end{align}
we have 
\begin{align}
\frac{\dd \phi_y}{\dd k_y}-i\varphi_-^{1\dagger}(\vec{k})\partial_{k_y}\varphi^1_-(\vec{k})=i \varphi_-^{2\dagger}(\vec{k})\partial_{k_y}\varphi^2_-(\vec{k}).
\end{align}
With the above formulas, we can obtain the 2D Zak-phases:
\begin{align}
\Omega^{2D}_x&=\frac{1}{2\pi}\int_{\text{BZ}}\,\Tr[\mathcal{A}_{x}(\vec{k})]\,\dd^2\vec{k}=\frac{1}{2\pi}\int dk_x\,\frac{\dd \phi_x}{\dd k_x}=\mathcal{N}_x,\\
\Omega^{2D}_y&=\frac{1}{2\pi}\int_{\text{BZ}}\,\Tr[\mathcal{A}_{y}(\vec{k})]\,\dd^2\vec{k}=\frac{1}{2\pi}\int dk_y\,\frac{\dd \phi_y}{\dd k_y}=\mathcal{N}_y.
\end{align}
The corresponding dipole moments along $x$ and $y$ directions are trivial since
\begin{align}
p_x&=\frac{\Omega^{2D}_x}{2\pi}\bmod 1=\mathcal{N}_x\bmod 1=0,\\
p_y&=\frac{\Omega^{2D}_y}{2\pi}\bmod 1=\mathcal{N}_y\bmod 1=0.
\end{align}

\section{Wannier-sector Polarization and Bulk Quadrupole Moment In 2D}
\subsection{Nested Wilson loop formalism}
To be well-defined, the topological quadrupole defined in Ref.~\cite{multipole_PRB_2017,multipole_science2017} applies only for a system with zero bulk Chern number, a vanishing dipole moment, and a gap in the Wannier-center bands. The last requirement originates from a quadrupole being constructed from two separated dipoles. Another crucial point for the generalized BBH model is that the Wannier bands in the $x$ and $y$ directions are both gapped, which hints at a separation of the two dipoles along the direction perpendicular to their aligned direction and thus yields the non-vanishing quadrupole.
  
Using the $j-$th Wannier band $v_j^x(k_y)$ with Wannier center $\nu_x^j(k_y)$ for the Wilson loop $\mathcal{W}_{x\vec{k}}$, we obtain the following projected-position eigenstates on the occupied bands at momentum $k_y$:
 \begin{align}
 \ket{\Psi^j_{R_x,k_y}}&=\sqrt{\frac{\Delta_{k_x}}{2\pi}}\sum_{n=1}^{N_\text{occ}}\sum_{k_x}[v_j^x(\vec{k})]_ne^{-i k_xR_x}|u_n(\vec{k})\rangle,
 \end{align}
which satisfies
\begin{align}
\hat{P}_{\text{occ}}\hat{g}_x \hat{P}_{\text{occ}}\ket{\Psi^j_{R_x,k_y}}=e^{i \Delta_{k_x}(\nu^j(k_y)+R_x)}\ket{\Psi^j_{R_x,k_y}},
\end{align}
where
\begin{align}
\hat{g}_x&=e^{i \Delta_{k_x}\mathtt{x}}=\sum_{\vec{k}, \alpha} \ket{\vec{k}+\vec{\Delta}_{k_x},\alpha}\bra{\vec{k}, \alpha},\\
\hat{P}_{\text{occ}}&=\sum_{n=1}^{N_{\text{occ}}}\ket{u_n(\vec{k})}\bra{u_n(\vec{k})},
\end{align}
and $N_{\text{occ}}$ is the number of the occupied bands. Note that we use the Dirac brackets to denote the state vectors and the symbols without brackets to denote their matrices representations. 

The Wannier sector is defined as the set that includes all the Wannier bands for the Wilson loop $\mathcal{W}_{\mu\vec{k}}$ $(\mu=x,y)$ with the same signs of the Wannier centers. We denote the Wannier sector as $\omega_\mu^s$, $s=\pm$. For every sector $\omega_\mu^s$ ($\mu=x,y$), there is polarization along the $\bar{\mu}$ ($=y,x$) direction, contributing to the quadrupole moment. Thus, to define the Wannier-sector polarization along $y$, we project the position operator $\hat{y}$ onto the Wannier sector $\omega_x^s$, $s=\pm$. The projected position operator $\hat{y}$ can be expressed as follows
\begin{align}
P_{\omega_x^s}g_yP_{\omega_x^s}=P_{\omega_x^s}\sum_{\vec{k}, \alpha} \ket{\vec{k}+\vec{\Delta}_{k_y},\alpha}\bra{\vec{k}, \alpha}P_{\omega_x^s},
\end{align}
where the projector for the sector below or above the Wannier gap is 
\begin{align}
P_{\omega_x^s}&=\sum_{j\in\omega_x^s}\sum_{n_1,n_2=1}^{N_\text{occ}}\sum_{\vec{k}}\ket{u_{n_2}(\vec{k})}\bra{ u_{n_1}(\vec{k})}[v_j^{x}(\vec{k})]_{n_2}[v_j^{x}(\vec{k})]_{n_1}^\ast.\label{sector_projection}
\end{align}
To simplify the notation, we will now use a special set of basis vectors called the Wannier-band basis,
\begin{align}
\ket{w_j^x(\vec{k})}&=\sum_n^{N_\text{occ}} \ket{u_n(\vec{k}) }[v_j^x(\vec{k})]_n. 
\end{align}
Therefore, the projected position operator reduces to 
\begin{align}
P_{\omega_x^s}g_yP_{\omega_x^s}=\sum_{j_1,j_2\in\omega_x^s}\sum_{\vec{k}}\ket{w^{x}_{j_1}(\vec{k}+\vec{\Delta}_{k_y})}\bra{ w^{x}_{j_2}(\vec{k})}(w_{j_1}^{x\dagger}(\vec{k}+\vec{\Delta}_{k_y})w_{j_2}^{x}(\vec{k})).
\end{align}
The eigenvalues of the sector-position operator $P_{\nu_x^s}g_yP_{\nu_x^s}$ give us the Wannier centers for the selected Wannier bands; we can diagonalize it by calculating the Wilson loop as
\begin{align}
\mathcal{W}^{\omega_x^{s}}_{y,\vec{k}}=\lim_{\Delta_{k_y}\to 0}F^{\omega_x^{s}}_{y,\vec{k}+(N_y-1)\vec{\Delta}_{k_y}}F^{\omega_x^{s}}_{y,\vec{k}+(N_y-2)\vec{\Delta}_{k_y}}...F^{\omega_x^{s}}_{y,\vec{k}+\vec{\Delta}_{k_y}}F^{\omega_x^{s}}_{y,\vec{k}},
\end{align}
where $[F^{\omega_x^{s}}_{y,\vec{k}}]_{j_1j_2}= w_{j_1}^{x\dagger}(\vec{k}+\vec{\Delta}_{k_y})w_{j_2}^{x}(\vec{k})$, $j_1,j_2\in\omega_x^s$.
The repeated indices imply the summation over all Wannier bands of the selected sector in the above definition. Usually, this Wilson loop, constructed from the Wannier-band basis, is usually called the ``nested Wilson loop". Since the nested-Wilson loop is unitary, the eigenvalues of $\mathcal{W}^{\omega_x{s}}_{k_y}$ are again phases:
\begin{align}
\mathcal{W}^{\omega_x^s}_{y,\vec{k}} \,v^{y}_{\omega_x^s,m}(\vec{k})=e^{i 2\pi\nu_y^{\omega_x^s,m}(k_x)}\, v^{y}_{\omega_x^s,m}(\vec{k}).
\end{align}

Finally, we can introduce the Wannier-sector polarization at $k_x$ as the summation of the Wannier-sector centers $\nu_y^{x,j}(k_x)$,
\begin{align}
p^{s}_y(k_x)=\sum_{m}\nu_y^{\omega_x^s,m}(k_x)\bmod 1.
\end{align}
The total polarization in the Wannier sector is 
\begin{align}
p^{s}_y=\frac{\Delta_{k_x}}{2\pi}\sum_{k_x}p^{s}_y(k_x)=\frac{1}{(2\pi)^2}\int_{\text{BZ}}\,\Tr[\mathcal{A}^{xs}_{y}(\vec{k})]\,\dd^2\vec{k}\bmod 1,
\end{align}
where $[\mathcal{A}^{xs}_{y}(\vec{k})]$ is the non-Abelian Berry connection for the Wannier bases of the Wannier sector $\omega_x^s$ and the trace $\Tr$ denotes the summation over all Wannier bands in this sector. By the same derivation, we can have the sector polarization in $x$ direction as
\begin{align}
p^{s}_x=\frac{\Delta_{k_y}}{2\pi}\sum_{k_y}p^{s}_x(k_y)=\frac{1}{(2\pi)^2}\int_{\text{BZ}}\,\Tr[\mathcal{A}^{ys}_{x}(\vec{k})]\,\dd^2\vec{k}\bmod 1.
\end{align}
The definition of the quadrupole moment is 
\begin{align}
q_{xy}=\sum_{s}p^{s}_xp^{s}_y.
\end{align}

\subsection{Quadrupole moment of the generalized BBH model}
In this part, we give the exact calculation of the quadrupole moment of the 2D generalized BBH model. 

Applying Eq.~(\ref{Psin}) in the expression for the Berry connection, we have
 \begin{align}
 \mathcal{A}_x(\vec{k})&=\left(\begin{array}{cc}
{i \matrixel{\psi_-^x(k_x)}{\partial_{k_x}}{\psi_-^x(k_x)}+i \matrixel{\varphi_-^1(\vec{k})}{\partial_{k_x}}{\varphi_-^1(\vec{k})}}&{i \matrixel{\psi_+^x(k_x)}{\partial_{k_x}}{\psi_-^x(k_x)}\,\varphi^1_-(\vec{k})^\dagger\varphi_-^2(\vec{k})} \\
{i \matrixel{\psi_-^x(k_x)}{\partial_{k_x}}{\psi_+^x(k_x)}\,\varphi^2_-(\vec{k})^\dagger\varphi_-^1(\vec{k})}&{i \matrixel{\psi_+^x(k_x)}{\partial_{k_x}}{\psi_+^x(k_x)}+i \matrixel{\varphi_-^2(\vec{k})}{\partial_{k_x}}{\varphi_-^2(\vec{k})}}
\end{array}\right),\\
 \mathcal{A}_y(\vec{k})&=\left(\begin{array}{cc}
{i \matrixel{\varphi_-^1(\vec{k})}{\partial_{k_y}}{\varphi_-^1(\vec{k})}}&{0} \\
{0}&{i \matrixel{\varphi_-^2(\vec{k})}{\partial_{k_y}}{\varphi_-^2(\vec{k})}}
\end{array}\right).
 \end{align}
Next, we express the Berry connections' elements in spherical coordinates. The derivation relies on the following relations
 \begin{align}
i \matrixel{\psi_\pm^x(k_x)}{\partial_{k_x}}{\psi_\pm^x(k_x)}&=i \matrixel{\psi_\pm^x(k_x)}{\partial_{k_x}}{\psi_\mp^x(k_x)}=\frac{1}{2}\frac{\dd \phi_x}{\dd k_x},\\
  i \matrixel{\varphi_-^1(\vec{k})}{\partial_{k_x}}{\varphi_-^1(\vec{k})}&=i \matrixel{\varphi_-^2(\vec{k})}{\partial_{k_x}}{\varphi_-^2(\vec{k})}=0,\\
 i \matrixel{\varphi_-^1(\vec{k})}{\partial_{k_y}}{\varphi_-^1(\vec{k})}&=\frac{\dd \phi_y}{\dd k_y}\sin{^2\theta_{xy}},\\
 i \matrixel{\varphi_-^2(\vec{k})}{\partial_{k_y}}{\varphi_-^2(\vec{k})}&=\frac{\dd \phi_y}{\dd k_y}\cos{^2\theta_{xy}},\\
\varphi_-^1(\vec{k})^\dagger\varphi_-^2(\vec{k})&=(\varphi_-^2(\vec{k})^\dagger\varphi_-^1(\vec{k}))^{\ast}=\sin{\theta_{xy}}\,.
 \end{align}
Thus, we obtain the following concrete expression for the Berry connections:
 \begin{align}
  \mathcal{A}_{x}(\vec{k})&=i \matrixel{\psi^x_-(k_x)}{\partial_{k_x}}{\psi^x_-(k_x)} s_0+\frac{1}{2}\sin{\theta_{xy}(\vec{k})}\frac{\dd \phi_x}{\dd k_x}s_1\nonumber\\&=\frac{1}{2}\frac{\dd \phi_x}{\dd k_x} \left(s_0+\sin{\theta_{xy}(\vec{k})}s_1\right),\\
    \mathcal{A}_{y}(\vec{k})&=i \matrixel{\varphi^1_-(\vec{k})}{\partial_{k_y}}{\varphi^1_-(\vec{k})} s_3=\frac{1}{2}\frac{\dd \phi_y}{\dd k_y}\left(s_0-\cos{\theta_{xy}}(\vec{k}) s_3\right).
  \end{align}
  By multiplying the exponent of these non-Abelian connections in path order, we determine the Wilson loops along the $x$ and $y$ axis as
  \begin{align}
  \mathcal{W}_{x,\vec{k}}(k_y)&= \exp{i \int_{k_x}^{k_x+2\pi}d\phi_x(k'_x)\frac{s_0+\sin{\theta_{xy}(k'_x,k_y)}s_1}{2}},\\
  \mathcal{W}_{y,\vec{k}}(k_x)&=\exp{i \int_{k_y}^{k_y+2\pi}d\phi_y(k'_y)\frac{s_0-\cos{\theta_{xy}(k_x,k'_y)s_3}}{2}}.
  \end{align}
  The above exact expressions for the Wilson loops allow us to directly solve the Wannier centers and the corresponding Wannier bases.  The Wilson loop along the $y$ direction has been defined to be diagonal,
  \begin{align}
      \nu^\pm_y(k_x)=\int_0^{2\pi}\frac{\dd k_y}{4\pi}\frac{\dd \phi_y}{\dd k_y}\left(1\pm\cos{\theta_{xy}(k_x,k_y)}\right).
  \end{align}
 For the Wilson loop $ \mathcal{W}_{x,\vec{k}}(k_y)$ along the $k_x$ axis through the point $\vec{k}$, we have
  \begin{align}
  \nu^\pm_x(k_y)&=\int_0^{2\pi}\frac{\dd k_x}{4\pi}\frac{\dd \phi_x}{\dd k_x}(1\pm\sin{\theta_{xy}(k_x,k_y)}).
  \end{align}
  Then the corresponding Wannier gaps are
  \begin{align}
    \Delta\nu_x(k_y)&=\int_0^{2\pi}\frac{\dd k_x}{2\pi}\frac{\dd \phi_x}{\dd k_x}\sin{\theta_{xy}(k_x,k_y)}=\int_0^{2\pi}\frac{\dd k_x}{2\pi}\frac{\dd \phi_x}{\dd k_x}\frac{|\vec{b}_y(k_y)|}{\sqrt{\vec{b}_x^2(k_x)+\vec{b}^2_y(k_y)}},\\
   \Delta\nu_y(k_x)&=\int_0^{2\pi}\frac{\dd k_y}{2\pi}\frac{\dd \phi_y}{\dd k_y}\cos{\theta_{xy}(k_x,k_y)}=\int_0^{2\pi}\frac{\dd k_y}{2\pi}\frac{\dd \phi_y}{\dd k_y}\frac{|\vec{b}_x(k_x)|}{\sqrt{\vec{b}_x^2(k_x)+\vec{b}^2_y(k_y)}}.
  \end{align}
 To calculate the Wannier-sector polarizations, we have to construct the Wannier-band bases:
 \begin{align}
 w_{j}^{\mu}(\boldsymbol{k})=\left(\Psi_{1}(\boldsymbol{k}), \Psi_{2}(\boldsymbol{k})\right) \cdot v_{j}^{\mu}(\boldsymbol{k}),
 \end{align}
  where $j$ corresponds to the $j-$th Wannier band and the Wannier sector $\omega_\mu^s$, $s=j=+,-$ since every sector only contains one Wannier band for the generalized BBH model.
  
 By solving the eigenstates of the Wannier loops in the energy representation, we obtain the following bases at the base point $\vec{k}_0$
 \begin{align}
     w^x_{j}(\vec{k}_0)&=\frac{1}{\sqrt{2}}\left(\Psi_1(\vec{k}_0)+j \Psi_2(\vec{k}_0)\right),\\
w^y_{1}(\vec{k}_0)&=\Psi_2(\vec{k}_0),\\
  w^y_{2}(\vec{k}_0)&=\Psi_1(\vec{k}_0).
 \end{align}
To consistently define the full Wannier basis, we maintain the gauge we choose in Eq.~(\ref{wannier_gauge}), which adds a $\vec{k}$-dependent phase to the bases. Recalling Eq.~(\ref{sector_projection}), we see that these phases cancel in the projector, so we can alternatively choose a simple family of Wannier bases to define the sector polarization as 
 \begin{align}
w^x_{j}(\vec{k})&=\frac{1}{\sqrt{2}}\left(\Psi_1(\vec{k})+j \Psi_2(\vec{k})\right),\\
w^y_{1}(\vec{k})&=\Psi_2(\vec{k}),\\
  w^y_{2}(\vec{k})&=\Psi_1(\vec{k}).
 \end{align}
 at any $\vec{k}$. We then use the above expressions to calculate the Berry connection for the Wannier basis,
\begin{align}
  [\mathcal{A}^{xs}_{y}(\vec{k})]_{jj}&=i \matrixel{ w_j^x(\vec{k})}{\partial_{k_y}}{w_j^x(\vec{k})}\nonumber\\
    &=\frac{1}{2}\Tr[\mathcal{A}_y(\vec{k})]+\frac{1}{2}\frac{\dd \phi_y}{\dd k_y}-j\,\text{Re}(e^{i \phi_y}[\mathcal{A}_y(\vec{k})]_{21})\nonumber\\
    &=\frac{1}{2}\frac{\dd \phi_y}{\dd k_y},\\
   [\mathcal{A}^{ys}_{x}(\vec{k})]_{jj}&=i \matrixel{ w_j^y(\vec{k})}{\partial_{k_x}}{w_j^y(\vec{k})}\nonumber\\
    &=[\mathcal{A}_x(\vec{k})]_{11}=[\mathcal{A}_x(\vec{k})]_{22}=\frac{1}{2}\frac{\dd \phi_x}{\dd k_x}.
    \end{align}
We integrate this to obtain the Wannier-sector polarizations as follows
  \begin{align}
p_x^{s}&=\frac{\mathcal{N}_x}{2}\bmod 1,\label{sec_px}\\
p_y^{s}&=\frac{\mathcal{N}_y}{2}\bmod 1\label{sec_py}.
\end{align}
Finally, the off-diagonal quadrupole moment as defined in Refs.~\cite{multipole_PRB_2017,multipole_science2017} is the multiplication of the two sector-polarizations
\begin{align}
q_{xy}=\sum_{s=\pm}p_x^{s}p_y^{s}=\frac{\mathcal{N}_x\mathcal{N}_y}{2}\bmod 1.
\end{align}

\section{Generalized BBH model with long-range hopping}
\begin{figure}
\centering
\includegraphics[width=0.35\textwidth]{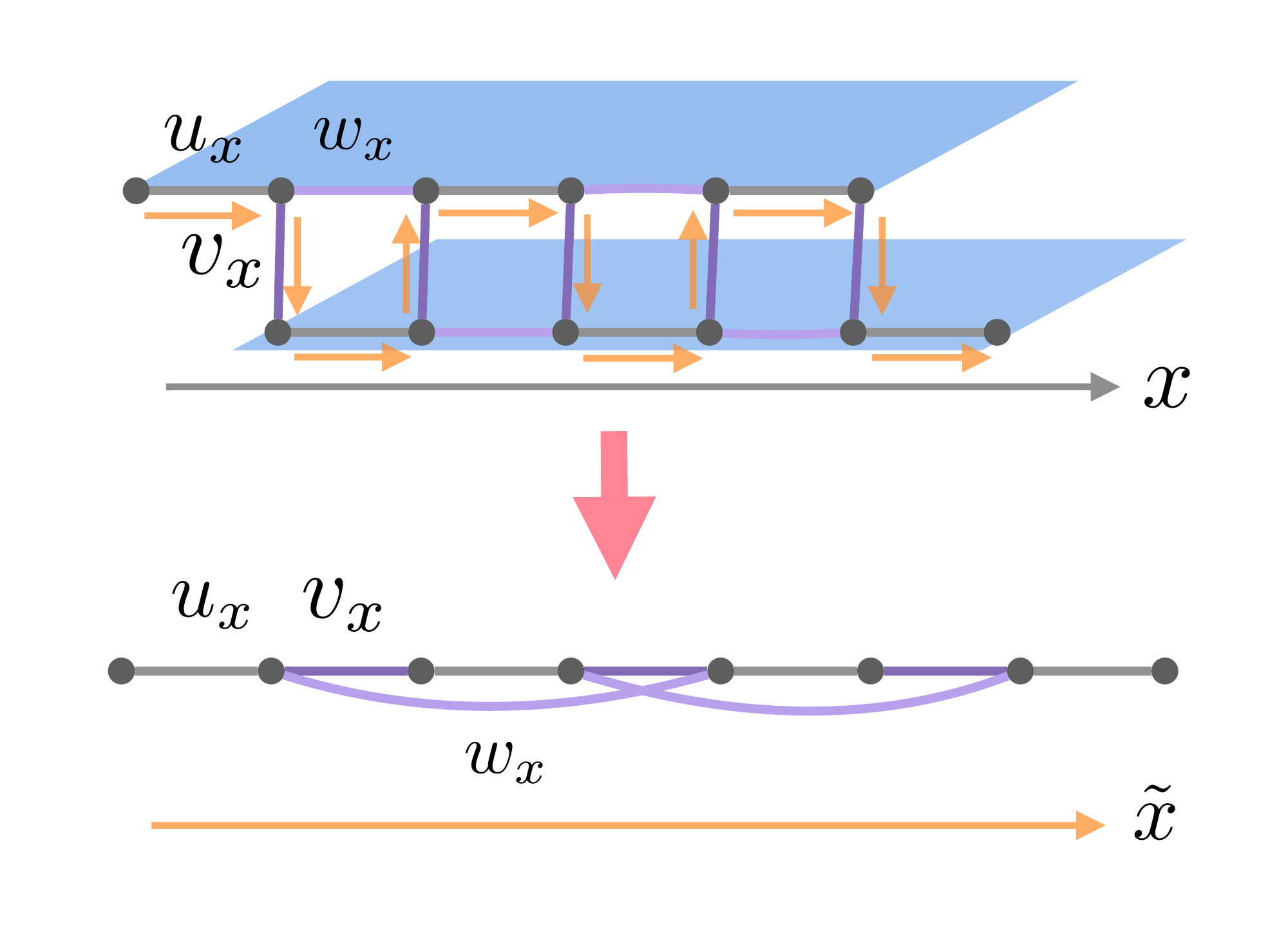}
\caption{Coupling of two BBH lattice layers with a stacking fault or AB stacking pattern. }\label{twolayer}
\end{figure}

In the main text, we propose a generalized BBH model with long-range hopping, whose Bloch Hamiltonian is
\begin{align}
\mathcal{H}(k_x,k_y)&=\vec{b}_x(k_x)\cdot\vec{\sigma}\otimes\tau_3 +\sigma_0\otimes \vec{b}_y(k_y)\cdot\vec{\tau}\label{longH},\\
\vec{b}_x(k_x)&=(u_x+v_x\cos{k_x}+w_x\cos{(2k_x)},v_x\sin{k_x}+w_x\sin{(2k_x)},0)\label{vectorbx}\\
\vec{b}_y(k_y)&=(u_y+v_y\cos{k_y},v_y\sin{k_y},0)\label{vectorby}.
\end{align}
To build such a model with tunable long-range hopping, we propose a two-layer system with a stacking fault, as shown schematically in Fig.~\ref{twolayer}. We can regard the two layers as a single-layer generalized BBH model, described by the Hamiltonian shown in Eq.~(\ref{longH}). The long-range hopping in the generalized model is the original nearest-neighbor hopping in a single layer. Interlayer hopping takes the role of the nearest-neighbor hopping term.

We now derive a method  to calculate the winding number $\mathcal{N}_x$ of the 1D two-band model described by the characteristic vector 
\begin{align}
\vec{b}_x(k_x)&=(u_x+v_x\cos{k_x}+w_x\cos{(2k_x)},v_x\sin{k_x}+w_x\sin{(2k_x)},0),
\end{align}
where $w_x$ corresponds to the long-range hopping. In an explicit matrix form, the Hamiltonian is
\begin{align}
h_x(k_x)=\left(\begin{array}{cc}
{0}&{u_x+v_xe^{-i k_x}+w_xe^{-2i k_x}} \\
{u_x+v_xe^{i k_x}+w_xe^{2i k_x}}&{0}
\end{array}\right).
\end{align}
As there is no term proportional to $\sigma_3$ because of the chiral symmetry, the eigenstates for this Hamiltonian can be written as
\begin{align}
\psi^x_\pm(k_x)=\frac{1}{\sqrt{2}}\left(\begin{array}{c}
{e^{-i [\tilde{\phi}_x(k_x)+k_x]} }\\
{\pm1}
\end{array}\right)=\left(\begin{array}{cc}
{e^{-i k_x}}&{0} \\
{0}&{1} 
\end{array}\right)\tilde{\psi}^x_\pm(k_x).
\end{align}
Here we define 
\begin{align}
\tilde{\psi}^x_\pm(k_x)=\frac{1}{\sqrt{2}}\left(\begin{array}{c}
{e^{-i \tilde{\phi}_x(k_x)} }\\
{\pm1}
\end{array}\right),
\end{align}
It can be shown that $\tilde{\psi}^x_\pm(k_x)$ are the eigenstates of the following Hamiltonian
\begin{align}
\tilde{h}_x(k_x)=\left(\begin{array}{cc}
{0}&{u_xe^{i k_x}+v_x+w_xe^{-i k_x}} \\
{u_xe^{-i k_x}+v_x+w_xe^{i k_x}}&{0}
\end{array}\right)=\tilde{\vec{b}}_x(k_x)\cdot\vec{\sigma},
\end{align}
where the characteristic vector is 
\begin{align}
\tilde{\vec{b}}_x(k_x)=(v_x+(u_x+w_x)\cos{k_x},(w_x-u_x)\sin{k_x},0).
\end{align}
We can see that the curve $\tilde{\mathcal{C}}:k_x\in[0,2\pi]\to \tilde{\vec{b}}_x(k_x)$ is actually an ellipse with semi-major axes along $k_x$ and $k_y$ as $|u_x+w_x|$ and $|w_x-u_x|$. Combining these properties and the winding direction determined by the signs in front of the cosine and sine in $\tilde{\vec{b}}_x(k_x)$, we can conclude the winding number of $\tilde{\mathcal{C}}$  is 
\begin{align}
\tilde{\mathcal{N}}_x= \begin{cases}0, & |u_x+w_x|<|v_x|  \\ \sign [(w_x+u_x)(w_x-u_x)], &|u_x+w_x|>|v_x|\end{cases}.\label{Nx_u}
\end{align}
This result has been shown previously, e.g., in Ref.~\cite{ExtendedSSH}. Applying the above results, we calculate the winding number for $\mathcal{C}:k_x\in[0,2\pi]\to \vec{b}_x(k_x)$ associated with the 1D Zak-Berry phase as
\begin{align}
\mathcal{N}_x=\frac{1}{2\pi}\int_0^{2\pi}\,\dd k_x\,i \matrixel{\psi_-^x(k_x)}{\partial_{k_x}}{\psi^x_-(k_x)}=1+\frac{1}{2\pi}\int_0^{2\pi}\,\dd k_x\,i \matrixel{\tilde{\psi}^x_-(k_x)}{\partial_{k_x}}{\tilde{\psi}^x_-(k_x)}=1+\tilde{\mathcal{N}}_x.
\end{align}
Finally, we obtain all possible values of  $\mathcal{N}_x$ with the corresponding conditions as
\begin{align}
\mathcal{N}_x= \begin{cases}1, & |u_x+w_x|<|v_x|  \\ \text{sign} [(w_x+u_x)(w_x-u_x)]+1 , &|u_x+w_x|>|v_x|\end{cases}.\label{Nx_u}
\end{align}

\section{Bulk Quadrupole Invariant}
We start from the quadrupole tensor definition in the Wannier representation
\begin{align}
\mathcal{N}_{\mu\nu}=\int_{\text{all}} \dd^2\vec{r} \,x_{\mu}x_{\nu}\, W_{n\vec{0}}^\dagger(\vec{r}) W_{n\vec{0}}(\vec{r}).
\end{align}
Using Bloch functions to expand the localized Wannier functions, we have
\begin{align}
W_{n\vec{R}}(\vec{r})&=\frac{1}{N}\sum_{k}e^{i\vec{k}\cdot(\vec{r}-\vec{R})}u_{n\vec{k}}(\vec{r}),\\
N&=\frac{(2\pi)^2}{\Delta_{k_x} \Delta_{k_y}}.
\end{align}
Now we use the Bloch function to express the quadrupole tensor as
\begin{align}
&\mathcal{N}_{\mu\nu}=-\frac{1}{N^2}\int_{\text{all}} \dd^2\vec{r} \sum_{k,k'}\sum_{n}\frac{\partial^2 e^{i(\vec{k}-\vec{k}')\cdot\vec{r}}}{\partial k_\mu\partial k_\nu}u_{nk'}^\dagger u_{nk}\nonumber\\
&=\frac{1}{N^2}\sum_{nkk'}\int_{\text{all}} \dd^2\vec{r} \frac{\partial}{\partial k_\mu}\left[-\frac{\partial e^{i(\vec{k}-\vec{k}')\cdot\vec{r}}}{\partial k_\nu}u_{nk'}^\dagger u_{nk}\right]+\frac{\partial}{\partial k_\nu}\left[e^{i(\vec{k}-\vec{k}')\cdot\vec{r}}u_{nk'}^\dagger \frac{\partial }{\partial k_\mu}u_{nk}\right]-\left[e^{i(\vec{k}-\vec{k}')\cdot\vec{r}}u_{nk'}^\dagger\frac{\partial}{\partial k_\nu}\frac{\partial }{\partial k_\mu} u_{nk}\right]\nonumber\\
&=\frac{1}{N^2}\sum_{nkk'}\int_{\text{all}} \dd^2\vec{r} \frac{\partial^2}{\partial k_\mu\partial k_\nu}\left[-e^{i(\vec{k}-\vec{k}')\cdot\vec{r}}u_{nk'}^\dagger u_{nk}\right]+\frac{\partial}{\partial k_\mu}\left[e^{i(\vec{k}-\vec{k}')\cdot\vec{r}}u_{nk'}^\dagger\frac{\partial u_{nk}}{\partial k_\nu} \right]+\frac{\partial}{\partial k_\nu}\left[e^{i(\vec{k}-\vec{k}')\cdot\vec{r}}u_{nk'}^\dagger \frac{\partial u_{nk}}{\partial k_\mu}\right]
\nonumber\\&\qquad\qquad
-e^{i(\vec{k}-\vec{k}')\cdot\vec{r}}u_{nk'}^\dagger\frac{\partial^2u_{nk}}{\partial k_\mu\partial k_\nu}\nonumber\\
&=\frac{1}{N}\sum_{nkk'}\int_{\text{cell}} \dd^2\vec{r} \frac{\partial^2}{\partial k_\mu\partial k_\nu}\left[-\delta_{kk'}u_{nk'}^\dagger u_{nk}\right]+\frac{\partial}{\partial k_\mu}\left[\delta_{kk'}u_{nk'}^\dagger\frac{\partial }{\partial k_\nu} u_{nk}\right]+\frac{\partial}{\partial k_\nu}\left[\delta_{kk'}u_{nk'}^\dagger \frac{\partial }{\partial k_\mu}u_{nk}\right]-\left[\delta_{kk'}u_{nk'}^\dagger\frac{\partial}{\partial k_\nu}\frac{\partial }{\partial k_\mu} u_{nk}\right]\nonumber\\
&=\frac{1}{N}\sum_{nk}\int_{\text{cell}} \dd^2\vec{r} \frac{\partial^2}{\partial k_\mu\partial k_\nu}\left[-u_{nk}^\dagger u_{nk}\right]+\frac{\partial}{\partial k_\mu}\left[u_{nk}^\dagger\frac{\partial }{\partial k_\nu} u_{nk}\right]+\frac{\partial}{\partial k_\nu}\left[u_{nk}^\dagger \frac{\partial }{\partial k_\mu}u_{nk}\right]-\left[u_{nk}^\dagger\frac{\partial}{\partial k_\nu}\frac{\partial }{\partial k_\mu} u_{nk}\right]\nonumber\\
&=\frac{1}{N}\sum_{k}\Tr\left[-i\frac{\partial}{\partial k_\nu}\mathcal{A}_\mu-i\frac{\partial}{\partial k_\mu}\mathcal{A}_\nu- u^\dagger_{nk}\frac{\partial^2}{\partial k_\mu\partial k_\nu}u_{nk}\right].\label{int_wannier}
\end{align}
In the above derivation, we have used the following formula
\begin{align}
\int_{\text{all}} \dd^2\vec{r} e^{i\vec{k}\cdot\vec{r}}f(\vec{r})=\sum_{R}\int_{\text{cell}}  \dd^2\vec{r} e^{i\vec{k}\cdot(\vec{r}+\vec{R})}f(\vec{r})=\sum_{R}e^{i\vec{k}\cdot\vec{R}}\int_{\text{cell}}  \dd^2\vec{r} e^{i\vec{k}\cdot\vec{r}}f(\vec{r})=N\delta_{k,0}\int_{\text{cell}}  \dd^2\vec{r} f(\vec{r}),
\end{align}
where $f(\vec{r}+\vec{R})=f(\vec{r})$ and $\vec{k}$ is discretized as $k_\mu=2\pi l_\mu/(N_\mu a_\mu)$ ($l_\mu$ is an integer) to make the wave function satisfy the periodic boundary condition. As $\mathcal{A}_\mu$ are all real, the term $-i\frac{\partial}{\partial k_\nu}\mathcal{A}_\mu-i\frac{\partial}{\partial k_\mu}\mathcal{A}_\nu$ in the integral of Eq.~(\ref{int_wannier}) should vanish.

In the continuum limit and with the localization condition $\langle \vec{R}_i|\vec{R}_j\rangle=\delta_{ij}$, we obtain the following topological quadrupole tensor as 
\begin{align}
\mathcal{N}_{\mu\nu}&=-\int \frac{\dd ^2\vec{k}}{(2\pi)^2}\Tr[\matrixel{ \Psi_{n}(\vec{k})}{\frac{\partial^2}{\partial k_\mu\partial k_\nu}}{\Psi_{n}(\vec{k})}],\label{quadrupole_invariant}
\end{align}
where the trace denotes the summation over all occupied bands. For the off-diagonal element $\mathcal{N}_{xy}$ which is our main focus in the main text, we have 
\begin{align}
\mathcal{N}_{xy}&=-\int \frac{\dd ^2\vec{k}}{(2\pi)^2}\Tr[\matrixel{ \Psi_{n}(\vec{k})}{\frac{\partial^2}{\partial k_x\partial k_y}}{\Psi_{n}(\vec{k})}]\nonumber\\
&=\int \frac{\dd ^2\vec{k}}{(2\pi)^2}(i\matrixel{\psi^x_+ }{\partial_{k_x}}{\psi^x_+})( i\matrixel{\varphi_-^1 }{\partial_{k_y}}{\varphi_-^1 })+(i\matrixel{\psi^x_-}{\partial_{k_x}}{\psi^x_-})( i\matrixel{\varphi_-^2 }{\partial_{k_y}}{\varphi_-^2 })\nonumber\\
&=\int \frac{\dd ^2\vec{k}}{(2\pi)^2}\frac{1}{2}\frac{\dd \phi_x}{\dd k_x}\frac{\dd \phi_y}{\dd k_y}\nonumber\\
&=\frac{\mathcal{N}_x\mathcal{N}_y}{2}.
\end{align}

We also calculate the two diagonal elements as
\begin{align}
\mathcal{N}_{xx}&=-\int \frac{\dd ^2\vec{k}}{(2\pi)^2} \Tr [\matrixel{ \Psi_{n}(\vec{k})}{\frac{\partial^2}{\partial k_x^2}}{\Psi_{n}(\vec{k})}]\nonumber\\
&=\int \frac{\dd ^2\vec{k}}{(2\pi)^2} \Tr [\matrixel{ \frac{\partial}{\partial k_x} \Psi_{n}(\vec{k})}{\frac{\partial}{\partial k_x}}{\Psi_{n}(\vec{k})}]\nonumber\\
&=\int \frac{\dd ^2\vec{k}}{(2\pi)^2} \Tr [\matrixel{ \frac{\partial}{\partial k_x} \psi_{n}^x}{\frac{\partial}{\partial k_x}}{\psi_{n}^x}]+\sum_{m=1,2}\matrixel{ \frac{\partial}{\partial k_x} \varphi_{-}^m}{\frac{\partial}{\partial k_x}}{\varphi_{-}^m}\nonumber\\
&=\int \frac{\dd ^2\vec{k}}{(2\pi)^2}[(\frac{\dd \phi_x}{\dd k_x})^2+\frac{1}{2}(\frac{\partial \theta_{xy}}{\partial k_x})^2],
\end{align}
and
\begin{align}
\mathcal{N}_{yy}&=\int \frac{\dd ^2\vec{k}}{(2\pi)^2} \Tr[-\matrixel{ \Psi_{n}(\vec{k})}{\frac{\partial^2}{\partial k_y^2}}{\Psi_{n}(\vec{k})}]\nonumber\\
&=\int \frac{\dd ^2\vec{k}}{(2\pi)^2} \Tr[\matrixel{ \frac{\partial}{\partial k_y} \Psi_{n}(\vec{k})}{\frac{\partial}{\partial k_y}}{\Psi_{n}(\vec{k})}]\nonumber\\
&=\int \frac{\dd ^2\vec{k}}{(2\pi)^2} \sum_{m=1,2}\matrixel{ \frac{\partial}{\partial k_y} \varphi_{-}^m}{\frac{\partial}{\partial k_y}}{\varphi_{-}^m}\nonumber\\
&=\int \frac{\dd ^2\vec{k}}{(2\pi)^2}[(\frac{\dd \phi_y}{\dd k_y})^2+\frac{1}{2}(\frac{\partial \theta_{xy}}{\partial k_y})^2].
\end{align}

We can see that the off-diagonal element $\mathcal{N}_{xy}$ is topologically invariant under continuous parameter deformation of the $x$- and $y$-direction constitutive chains without closing the Wannier gaps. 

\section{ Edge polarizations and edge-consistent gauge}
In Refs.~\cite{multipole_PRB_2017,multipole_PRB_2017,Changan2020}, it has been shown through numerical calculation of the hybrid Wannier functions in the BBH model that the fractional edge polarizations parallel to the edge are localized at the boundary in ribbon systems. Thus, the dipole moments of the edge states should capture all the edge polarizations when the thermodynamic limit is taken. Using this condition, we derive the exact edge polarizations for the generalized BBH model around the boundary.

First, we consider a ribbon along the $x$-axis described by the generalized BBH model, where the occupied $m_x-$th edge states along the up- or down-edge (labeled by U/D) are
\begin{align}
\Psi^\text{U/D}_{m_x}(k_x)=\psi^x_{-\sigma_{m_x}}(k_x)\otimes\varphi_{ \text{U/D}}^{\sigma_{m_x}},\quad {m_x}=1,..,\mathcal{N}_y,\label{edgex}
\end{align} 
where $h_x(k_x)\psi^x_{-\sigma_{m_x}}(k_x)=-\sigma_{m_x}|\vec{b}_x(k_x)|\,\psi^x_{-\sigma_{m_x}}(k_x)$, $h_y\varphi^{\sigma_{m_x}}_{\text{U/D}}=0$ and $\tau_3\varphi^{\sigma_{m_x}}_{\text{U/D}}=\sigma_{m_x}\varphi^{\sigma_{m_x}}_\text{U/D}$ ($\sigma_{m_x}=\pm1$). $\varphi^{\sigma_{m_x}}_{\text{U/D}}$ and $\sigma_{m_x}$ labels the chirality. If we apply the gauge in Eq.~(\ref{TwoBand}) for $\psi^x_{-\sigma_{m_x}}(k_x)$, then we obtain the edge polarization along $x$-axis as
\begin{align}
p^\text{edge}_x&=\sum_{m_x=1}^{\mathcal{N}_y}\frac{1}{2\pi }\int_0^{2\pi}dk_x\,i\matrixel{\Psi^\text{U/D}_{m_x}(k_x)}{\partial_{k_x}}{\Psi^\text{U/D}_{m_x}(k_x)}\bmod 1\nonumber\\
&=\frac{\mathcal{N}_x}{2}\mathcal{N}_y \bmod 1.
\end{align}
Similarly, the occupied $m_y-$th edge states for a ribbon along the $y$-axis are
\begin{align}
\Psi^\text{L/R}_{m_y}(k_x)=\varphi^{\tau_{m_y}}_{\text{L/R}}\otimes\psi_{-}^{y}(k_y),\quad {m_y}=1,..,\mathcal{N}_x,\label{edgey}
\end{align}
where $h_y(k_y)\psi^y_{-}(k_y)=-|\vec{b}_y(k_y)|\,\psi^y_{-}(k_y)$, $h_x\varphi^{\tau_{m_y}}_{\text{L/R}}=0$ and $\sigma_3\varphi^{\tau_{m_y}}_{\text{L/R}}=\tau_{m_y}\varphi^{\tau_{m_y}}_\text{L/R}$ ($\tau_{m_y}=\pm1$). L/R denotes the edge where the zero modes $\varphi^{\tau_{m_y}}_{\text{L/R}}$ localize in the $x$-direction and $\tau_{m_y}$ gives the chirality of it. The corresponding edge polarization parallel to the $y$-axis is
\begin{align}
p^\text{edge}_y&=\sum_{m_y=1}^{\mathcal{N}_x}\frac{1}{2\pi }\int_0^{2\pi}dk_y\,i\matrixel{\Psi^\text{L/R}_{m_y}(k_y)}{\partial_{k_y}}{\Psi^\text{L/R}_{m_y}(k_y)}\bmod 1\nonumber\\
&=\frac{\mathcal{N}_y}{2}\mathcal{N}_x \bmod 1.
\end{align}
These results are consistent with the typical BBH model with $p^\text{edge}_x=p^\text{edge}_y=1/2$ in the topological phase~\cite{multipole_PRB_2017,multipole_science2017}.

In the main text, we propose a specific gauge for the bulk and edge Wannier functions. To address the edge-consistent properties of this gauge, we study the bulk and edge Wannier functions of the central unit cell 
\begin{align}
|W^{\text{bulk}}_{n\vec{0}}\rangle&=(N_xN_y)^{-1}\sum_{\vec{R}\vec{k}\alpha_x\beta_y}e^{i\vec{k}\cdot\vec{R}}[\Psi_{n}(\vec{k})]_{\alpha_x,\beta_y}|\vec{R},\alpha_x,\beta_y\rangle,\\
|W^{\text{U/D}}_{m_x\vec{0}}\rangle&=N_x^{-1}\sum_{R_xk_x\alpha_x}e^{ik_xR_x}[\psi^x_{-\sigma_{m_x}}(k_x)]_{\alpha_x}|R_x,\alpha_x\rangle,\\
|W^{\text{L/R}}_{m_y\vec{0}}\rangle&=N_y^{-1}\sum_{R_yk_y\beta_y}e^{ik_yR_y}[\psi_{-}^{y}(k_y)]_{\beta_y}|R_y,\beta_y\rangle,
\end{align}
with their corresponding densities
\begin{align}
\rho^\text{bulk}_n(\vec{R},\alpha_x,\beta_y)&=|\langle\vec{R},\alpha_x,\beta_y|W^{\text{bulk}}_{n\vec{0}}\rangle|^2,\\
\rho^\text{U/D}_{m_x}(R_x,\alpha_x)&=|\langle R_x,\alpha_x|W^{\text{U/D}}_{m_x\vec{0}}\rangle|^2,\\
\rho^\text{L/R}_{m_y}(R_y,\beta_y)&=|\langle R_y,\beta_y|W^{\text{L/R}}_{m_y\vec{0}}\rangle|^2.
\end{align}
\begin{figure}
\centering
\includegraphics[width=0.7\textwidth]{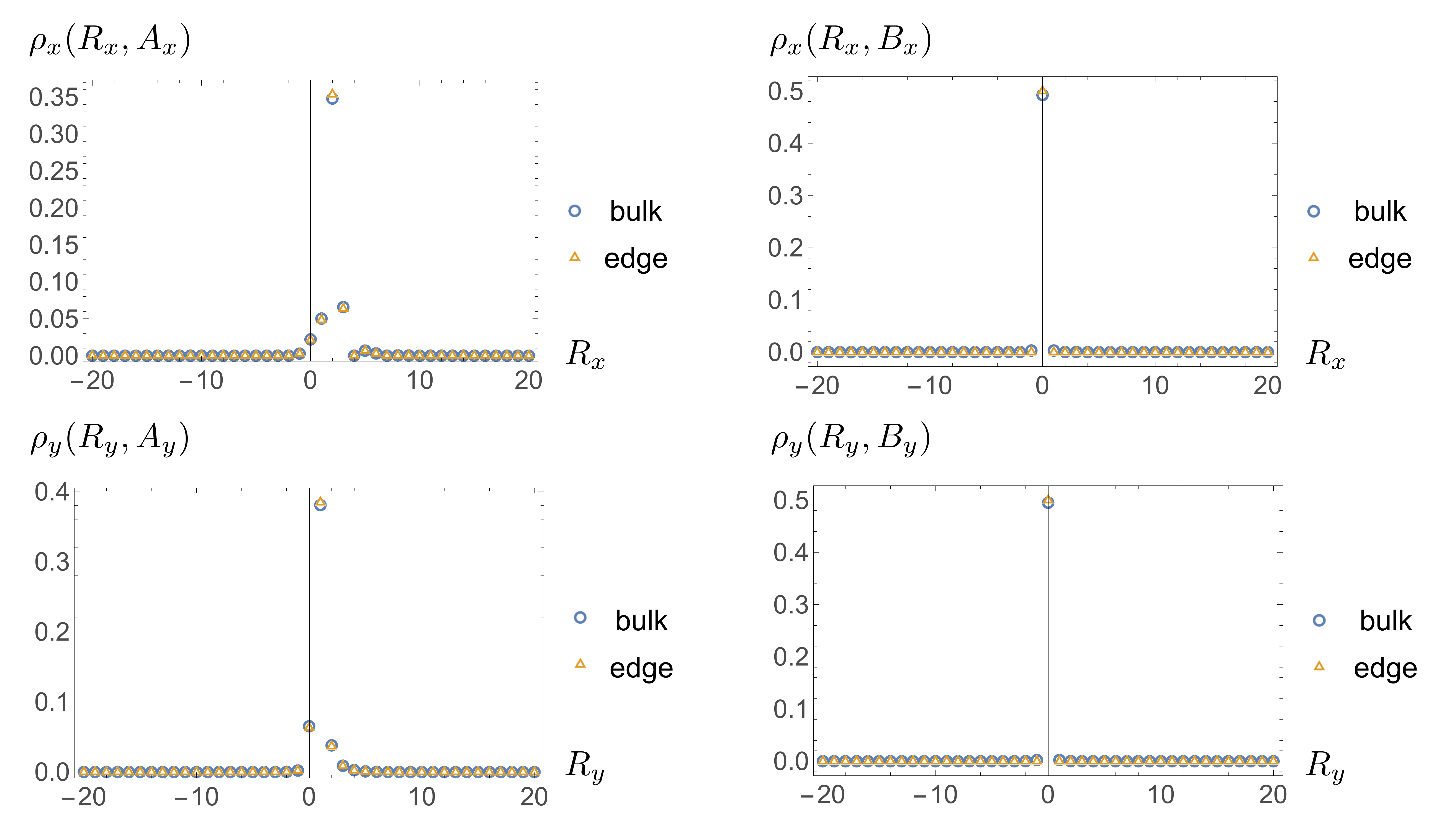}\\
\caption{The reduced probability density of the bulk and edge Wannier functions at the corresponding center unit cells of the model with long-range hopping described in Eq.~(\ref{longH}). The parameters used are $u_x=u_y=1$, $v_x=v_y=3/2$, and $w_x=2$, which are chosen to generate the topological phase. We can see the 2D bulk Wannier functions and the 1D edge states have a common density distribution which shows the gauge we choose for the bulk is edge-consistent.}\label{edge_gauge}
\end{figure}
For an edge-consistent gauge, the reduced density of the center Wannier functions is required to satisfy as we presume in the discussion section of the main text \begin{align}
\frac{1}{N_\text{occ}}\sum_{n=1}^{N_\text{occ}}\sum_{R_y,\beta_y}\rho^\text{bulk}_n(\vec{R},\alpha_x,\beta_y)=\frac{1}{\mathcal{N}_y}\sum_{m_x=1}^{\mathcal{N}_y}\rho^\text{U/D}_{m_x}(R_x,\alpha_x)=\rho_x(R_x,\alpha_x),\\
\frac{1}{N_\text{occ}}\sum_{n=1}^{N_\text{occ}}\sum_{R_x,\alpha_x}\rho^\text{bulk}_n(\vec{R},\alpha_x,\beta_y)=\frac{1}{\mathcal{N}_x}\sum_{m_y=1}^{\mathcal{N}_x}\rho^\text{L/R}_{m_y}(R_y,\beta_y)=\rho_y(R_y,\beta_y).
\end{align}
To check whether the gauge we use in the main text for the generalized BBH model satisfies the above conditions, we use the numerical calculation for the Wannier functions of the system described in Eq.~(\ref{longH}) with $u_x=u_y=1$, $v_x=v_y=3/2$ and $w_x=2$ and show the consistency of the reduced density distribution between the edge and bulk in Fig.~\ref{edge_gauge}.

\section{Gauge transformation of the Quadrupole invariant}
The quadrupole invariant we defined in Eq. (\ref{quadrupole_invariant}) can be expressed by the quantum geometry metric~\cite{Resta2011,geometry_metric2000,Watanabe2019} as
\begin{align}
\mathcal{N}_{\mu\nu}&=\int_{\text{FBZ}} \frac{\dd ^2\vec{k}}{(2\pi)^2}g_{\mu\nu}(\vec{k})+\text{Tr}[\mathcal{A}_\mu\mathcal{A}_\nu]\nonumber\\
g_{\mu\nu}(\vec{k})&=\operatorname{Re} \sum_{n=1}^{N_{\text{occ}}} (\partial_{k_{\mu}} \Psi_{n }(\vec{k}))^\dagger \partial_{k_{\nu}} \Psi_{m }(\vec{k})-\sum_{n=1}^{N_{\text{occ}}} \sum_{m=1}^{N_{\text{occ}}} (\partial_{k_{\mu}} \Psi_{n }(\vec{k}))^\dagger  \Psi_{m }(\vec{k})\,\Psi_{m }^\dagger (\vec{k}) \partial_{k_{\nu}} \Psi_{n}( \vec{k}).
\end{align}
Note that the integral of the quantum geometry metric is gauge invariant and equal to $\int_{\text{FBZ}} \dd ^2\vec{k}\,g_{\mu\nu}(\vec{k})/(2\pi)^2=0$. Applying the solutions of $\{\Psi_{1}(\vec{k}),\Psi_{2}(\vec{k})\}$, we find the off-diagonal quadrupole moment as
\begin{align}
\mathcal{N}_{xy}&=\int_{\text{FBZ}} \frac{\dd ^2\vec{k}}{(2\pi)^2}(g_{xy}(\vec{k})+\text{Tr}[\mathcal{A}_x\mathcal{A}_y])\nonumber\\
&=\int_{\text{FBZ}} \frac{\dd ^2\vec{k}}{(2\pi)^2}\text{Tr}[\mathcal{A}_x\mathcal{A}_y].
\end{align}
Under the following $U(2)$ gauge transformation 
\begin{align}
\left(\begin{array}{cc} \Psi_1(\vec{k})&\Psi_2(\vec{k})\end{array}\right)\rightarrow\left(\begin{array}{cc} \Psi_1(\vec{k})&\Psi_2(\vec{k})\end{array}\right)U,
\end{align}
the non-Abelian Berry connections $\mathcal{A}_\mu$ transform as
\begin{align}
\mathcal{A}_\mu\to U^\dag \mathcal{A}_\mu U+U^\dag i\partial_{k_\mu} U
\end{align}
and the corresponding transformation for the quadrupole moment is
\begin{align}
\mathcal{N}_{xy}\to\mathcal{N}_{xy}+\frac{1}{(2 \pi)^{2}} \int_{\mathrm{FBZ}} d \vec{k} \,\text{Tr}\{\mathcal{A}_x (-iU\partial_{k_y} U^\dag)+\mathcal{A}_y (-iU\partial_{k_x} U^\dag)+(\partial_{k_x}U^\dag)(\partial_{k_y}U)\}.\label{Nxy_gauge}
\end{align}
Since the edge states are all quasi-1D states and should have a common gauge for different bands in Eqs. (\ref{edgex}) and (\ref{edgey}), the physical gauge transformation for edge in $\mu-$direction can only be a phase factor 
\begin{align}
U_\mu(k_\mu)=e^{-i\Theta_\mu(k_\mu)}.
\end{align}
Following the derivation in Eq. (\ref{1D_p_gauge}), we can obtain the corresponding transformation for the edge polarizations are
\begin{align}
p^\text{edge}_\mu&\rightarrow\frac{\mathcal{N}_\mu}{2}\mathcal{N}_{\tilde{\mu}}+Z_\mu\mathcal{N}_{\tilde{\mu}}\bmod 1=p^\text{edge}_\mu,\\
Z_\mu&=\frac{\Theta_\mu(2\pi)-\Theta_\mu(0)}{2\pi},
\end{align}
where $\bar{\mu}=y,x$ when $\mu=x,y$. 

With the relation between the corner charge $Q_c$, edge polarizations, and the bulk quadrupole, we presume that the physical gauge choice for the 2D bulk states should be restricted as 
\begin{align}
U(\vec{k})=U_x(k_x)U_y(k_y)=e^{-i(\Theta_x(k_x)+\Theta_y(k_y))}.
\end{align}
Applying this transformation to Eq. (\ref{Nxy_gauge}), we have
\begin{align}
\mathcal{N}_{xy}&\to\mathcal{N}_{xy}+\mathcal{N}_{x}Z_y+\mathcal{N}_{y}Z_x+2Z_xZ_y,
\end{align}
where $Z_\mu$ are integers for periodic $U(\vec{k})$. Note that the above transformation preserves the fractional quantization of $\mathcal{N}_{xy}$. 

Similarly, the Wannier-sector polarizations and the quadrupole moment change as follow with gauge transformation $U$:
\begin{align}
p_x^{s}&\rightarrow \frac{\mathcal{N}_x}{2}+Z_x\bmod 1\\
p_y^{s}&\rightarrow \frac{\mathcal{N}_y}{2}+Z_y\bmod 1\\
q_{xy}&\rightarrow 2(\frac{\mathcal{N}_x}{2}+Z_x)(\frac{\mathcal{N}_y}{2}+Z_y)\bmod 1=\mathcal{N}_{xy}+\mathcal{N}_{x}Z_y+\mathcal{N}_{y}Z_x+2Z_xZ_y\bmod 1.
\end{align}
Here we can see that the corresponding gauge transformation of $q_{xy}$ and $\mathcal{N}_{xy}$ does not break their relation $q_{xy}=\mathcal{N}_{xy}\bmod 1$, which hints the separable gauge transformation is physical.

\bibliography{BBH.bib}